# Chondrule formation in the outer disk from the primary three-dimensional chemical composition of CM chondrules

Poula Eyðbjørnsdóttir[1*], Anders Johansen[1], Elishevah van Kooten[1]

[1]Center for Star and Planet Formation, Globe Institute, University of Copenhagen, Øster Voldgade 5-7, 1350 Copenhagen, Denmark. * poula.eydbjornsdottir@sund.ku.dk

**Abstract**
Chondrules and their associated fine-grained rims (FGRs) record fundamental processes operating in the early protoplanetary disk, yet the relationship between chondrule chemistry, morphology, and matrix complementarity remains incompletely understood. Here we investigate the major, minor, and trace element compositions of 66 chondrules and associated FGRs from relatively unaltered CM carbonaceous chondrites Asuka 12236, Paris and Maribo in relation to their three-dimensional morphology, using a multi-analytical approach including femtosecond LA-ICP-MS and X-ray tomography. Our results show that CM chondrules record a systematic process of metal loss and evaporation of Si-rich mesostasis, driving initially CI-like precursor compositions toward more Mg- and Si-rich bulk compositions along the CI ratio line and toward increasingly Si-poor forsteritic mineral assemblages. GEMS-like materials in pristine CM matrices appear to mirror chondrule compositions and likely represent complementary condensates derived from evaporated Si-rich mesostasis. The dust accreted to chondrules is dominantly CI-like but incorporates ~14 wt.% complementary condensate material represented by chondritic amorphous silicates, reconciling the observed Mg/Si complementarity between chondrules and matrix with the preservation of primordial organics and presolar grains.
Morphological observations further reveal no significant sectioning bias in chondrule size or plane, consistent with CM chondrule populations being dominated by agglomerates of ~100 µm sized microspherules rather than larger primary melt droplets. Many chondrules display grape-bunch textures formed by welding of smaller primary chondrules with metal-rich or CI-like rims. This structure may explain the moderate volatile element plateau at ~0.3×CI observed for average CM chondrule compositions, reflecting incorporation of primary fine-grained rim material into these aggregates. We propose a "micro-chondrule-first" formation scenario in which localized heating events produced small molten droplets that subsequently accreted CI-like dust and ice, aggregated, and experienced limited in situ aqueous alteration. These observations place new constraints on chondrule formation mechanisms in the outer disk and highlight the importance of localized melting and aggregation processes.

## 1. Introduction

Chondritic meteorites represent the most primitive solids available for investigating the early evolution of the Solar System, having experienced relatively limited modification since their accretion in the protoplanetary disk. A defining component of chondrites is chondrules—once-molten, millimeter-sized spherules that formed throughout the disk's lifetime (Connelly et al., 2012; Bollard et al., 2017; Schrader et al., 2017; Budde et al., 2018) and in both the inner and outer disk reservoirs (Olsen et al., 2016; van Kooten et al., 2016; Schneider et al., 2020; Williams et al., 2020; van Kooten et al., 2020, 2021). As the principal constituents of chondritic parent bodies and key ingredients in pebble accretion models (Johansen et al., 2015; Garai et al., 2025), chondrules provide fundamental constraints on both disk-scale transport processes and planetesimal formation. Most chondrules are coated by fine-grained rims (FGRs), which consist of accreted dust subsequently compacted and cemented on chondrite parent bodies (Brearley, 1993; Huss et al., 2005).

Mighei-type (CM) chondrites constitute the most abundant group among carbonaceous chondrites in terrestrial meteorite collections and are widely regarded as representative of materials that formed in the outer regions of the protoplanetary disk (Scott, 2007). Their chondrules and associated FGRs therefore offer an exceptional opportunity to investigate the physicochemical and accretionary processes that govern the outer disk. CM chondrules are thought to be representative for chondrule formation in the outer disk (van Kooten et al., 2020) and could therefore shed light on fundamental questions including to whether chondrules and their FGRs formed locally and are genetically related (Hezel and Palme, 2010; Ebel et al., 2016; Hezel et al., 2018a) or if transport of these components is required (van Kooten et al., 2019, 2024a). However, despite decades of study, the origin and evolution of chondrules and their fine-grained rims remain actively debated. A major obstacle is the pervasive effects of secondary aqueous alteration in CM chondrites (Rubin et al., 2007; King et al., 2017), which can redistribute elements and obscure primary chemical signatures (Zanda et al., 2011, 2018; Alexander, 2019; van Kooten et al., 2019; Patzer et al., 2023; van Kooten et al., 2024a). In addition, most previous studies rely on two-dimensional sectioning or bulk chemical analyses, which inherently fail to capture the true three-dimensional geometry of chondrules and their rims (Hezel, 2007; Hezel and Kießwetter, 2010; Barosch et al., 2020).

Two-dimensional analytical approaches introduce significant sectioning bias, as randomly cut sections do not accurately represent chondrule size, shape, internal structure, or rim thickness (Metzler, 2018; Barosch et al., 2020). These biases propagate into interpretations of accretion textures, chemical zoning, and alteration histories. While elemental mapping techniques such as scanning electron microscopy (SEM)-based analyses and laser ablation inductively coupled plasma mass spectrometry (LA-ICP-MS) provide high spatial resolution, they are typically restricted to planar surfaces. Three-dimensional methods, such as X-ray computed tomography (XCT), offer a non-destructive means to visualize entire chondrules, their internal fabrics, and the true morphology of their fine-grained rims. However, XCT studies of CM chondrites remain limited, in part because overlapping X-ray attenuation coefficients of major silicate phases and the compositional similarity between rims and matrix complicate phase segmentation and quantitative analysis.

In this study, we overcome these limitations by integrating high-resolution XCT imaging with SEM and femtosecond LA-ICP-MS elemental mapping to investigate three of the most unaltered CM chondrites available: Paris, Maribo, and Asuka 12236 (Haack et al., 2012; Hewins et al., 2014; Kimura et al., 2020). By combining two-dimensional chemical data with three-dimensional structural information, we characterize the geometry and composition of chondrules and their fine-grained rims, assess chondrule-matrix complementarity and volatile element systematics, and quantitatively evaluate the extent to which sectioning bias has influenced previous interpretations. This integrated approach provides new constraints on the processes governing chondrule formation, rim accretion, and early Solar System accretion dynamics.

## 2. Materials and methods

### *2.1 Samples and preparation*

The three relatively unaltered CM chondrites with limited degrees of aqueous alteration (Rubin et al., 2007) – Maribo (CM2.7, Haack et al., 2011), Asuka 12236 (CM2.9, Kimura et al., 2020), and Paris (CM2.7-2.9, Hewins et al., 2014) – were used for this study. For Maribo, two 1×1 cm slices of 1 mm thick were cut dry from a hand sample from the Natural History Museum of Denmark using a diamond wire saw at the Centre for Star and Planet Formation (StarPlan). For Paris and Asuka 12236, two 2×2 cm thick sections of 4 mm thickness were loaned by the Natural History Museum of Paris and the Royal Belgian Institute of Natural Sciences, respectively. All samples were mounted in 1-inch epoxy mounts. While Paris and Asuka 12236 were polished at their respective home facilities, Maribo sections were polished at StarPlan using MQ water on SiC sheets (>5 µm grit size) and $Al_2O_3$ sheets (>0.1 µm grit size). Below, we provide a brief description of the samples based on observations from literature.

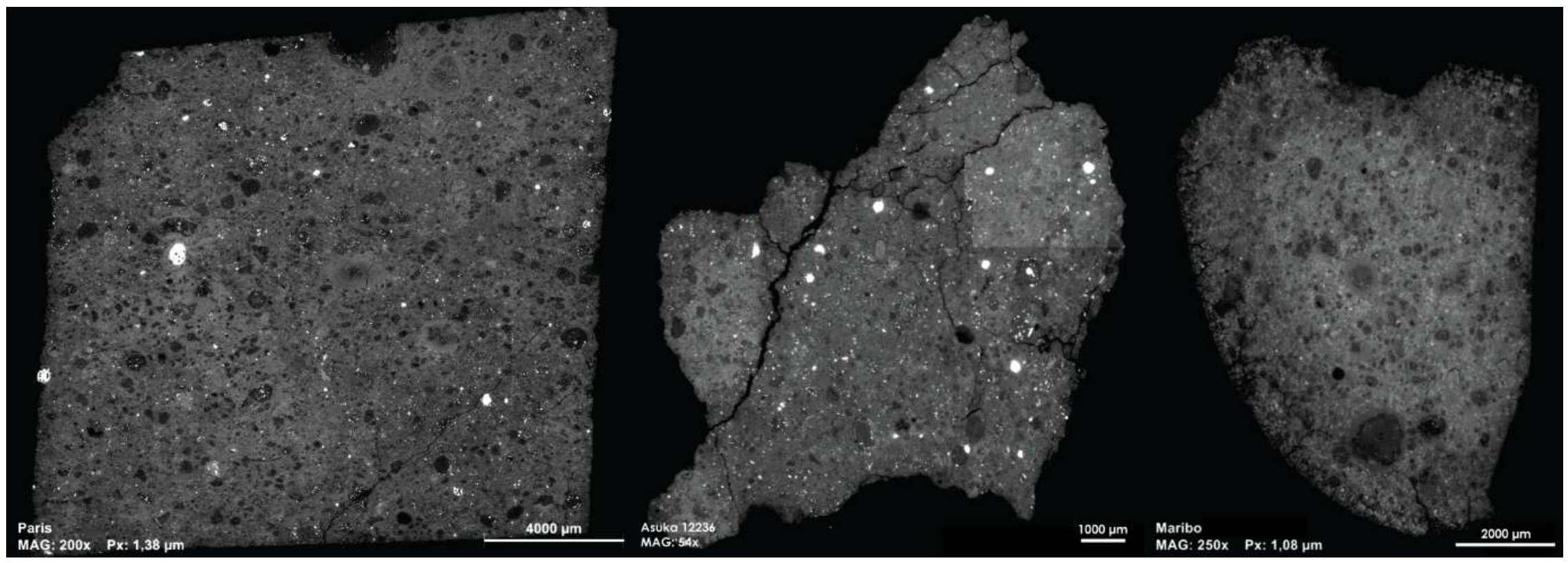

***Figure 1:*** *Back scattered electron mosaic images of CM chondrite sections used in this study. From left to right: Paris, Asuka 12236 and Maribo (section B). Note that the upper right corner of Asuka 12236 is brighter due to stitching effects during rendering of the BSE mosaic image.*

*2.1.1 Asuka 12236*

Based on its presolar grain inventory (Nittler et al., 2021), modal abundances, bulk mineralogical compositions (almost no secondary minerals) and its chondrule size distribution (Kimura et al., 2020), Asuka 12236 is described as one of the most unaltered CM chondrites in the meteorite collection. Infrared and Raman spectroscopy highlights the pristine nature of this chondrite through identification of a high abundance of anhydrous silicates and amorphous phases, even relative to Paris (Djouadi et al., 2025). In addition, the amino acid inventory and characteristics of the soluble organic matter suggest a lower degree of aqueous alteration for Asuka 12236 (Glavin et al., 2020; Serra et al., 2023). Focusing on the petrological features of this chondrite relevant for this study, estimated modal areas of Asuka 12236 components include 29% chondrules, 65% matrix, 4% refractory inclusions, 1% metal and 1% sulfides (Kimura et al., 2020). The average diameter of the chondrules is 290 µm and most chondrules are porphyritic (97.9%), with 92% of the chondrules being type I (FeO-poor; Kimura et al., 2020). Roughly half of the chondrules retain their primary mesostasis and tochilinite-cronstedtite-intergrowths (TCIs) are not observed within any of the chondrules (Kimura et al., 2020). These objects contain FeNi metal grains within and at their outer margins. Most chondrules are surrounded by FGRs. The high abundance of sulfur in the matrix, as well as the low abundance of phyllosilicates and TCIs, is a signature of the low degree of aqueous alteration experienced by Asuka 12236, rather than signifying devolatilization by secondary heating (Kimura et al., 2020).

*2.1.2 Paris*

Based on its mineralogical, petrological, and organic characteristics, the CM chondrite Paris is widely regarded as one of the least aqueously altered CM chondrites and serves as a reference material for pristine CM compositions (Hewins et al., 2014; Rubin, 2015). Detailed petrographic and mineralogical investigations show that Paris is heterogeneous at the thin-section scale, containing domains that span a range of aqueous alteration states, commonly estimated between CM2.7 and CM2.9 (Marrocchi et al., 2014; Hewins et al., 2014). Modal abundances and bulk mineralogical data indicate that Paris preserves a high abundance of anhydrous silicates in both chondrules and matrix, with secondary hydrous phases present but limited in extent. Infrared and Raman spectroscopy document the coexistence of crystalline anhydrous silicates and abundant amorphous material, consistent with incipient and spatially variable aqueous alteration (Djouadi et al., 2025). Compared to Asuka 12236, Paris contains higher abundances of phyllosilicates and rare tochilinite–cronstedtite intergrowths (TCIs), indicating a slightly greater degree of parent-body aqueous processing.
Organic inventories provide independent support for this interpretation. The distribution and characteristics of soluble organic matter and amino acids in Paris record aqueous alteration signatures that are weaker than in most CM chondrites but more developed than in Asuka 12236 (Glavin et al., 2010, 2020). Paris is composed predominantly of fine-grained matrix (~55 vol.%), with chondrules comprising <45 vol.% of the rock and minor abundances of refractory inclusions, metal, and sulfides (Marrocchi et al., 2014; Hewins et al., 2014). Chondrules have average diameters of ~250 μm and are dominated by porphyritic textures, with type I chondrules being the most abundant. Many chondrules retain partially preserved primary mesostasis, although localized replacement by phyllosilicates is observed, in contrast to the more extensive preservation of primary mesostasis and absence of TCIs in Asuka 12236. Overall, Paris experienced a low but measurable and spatially heterogeneous degree of aqueous alteration, placing it slightly more altered than Asuka 12236.

*2.1.3 Maribo*
In contrast to the other CM chondrites investigated in this study, Maribo is an observed fall that was recovered after exposure to approximately one month of winter weather conditions in Denmark. Consequently, it exhibits signs of terrestrial weathering at the bulk scale, including partial leaching of Na and K (Haack et al., 2012). Despite this overprint, multiple lines of evidence indicate that Maribo experienced only limited parent-body aqueous alteration. The organic inventory, nanoscale mineralogy, and hydrogen and nitrogen isotope systematics suggest a high degree of preservation of primordial organic matter inherited from the interstellar medium (van Kooten et al., 2018; Vollmer et al., 2020). Maribo is more matrix-rich relative to Paris and Asuka 12236, with <10 vol.% chondrules, typically <400 μm in diameter (Haack et al., 2012). It contains abundant FGRs surrounding chondrules, as well as tochilinite and cronstedtite grains. In contrast to Paris and Asuka 12336, large metal grains are not observed, but rare sulfides are present. In addition, Maribo does not contain an intra-chondrule matrix (i.e., the matrix between chondrules excluding the FGRs), but the matrix fraction is retained in the FGRs.

| | | Maribo | | |
|---|---|---|---|---|
| **Diameter** (μm) | **Polar** | **Midlatitude** | **Equatorial** | *Undefined* |
| < 300 | 8 | 3 | 1 | - |
| 300-500 | 3 | 2 | - | - |
| > 500 | 1 | 3 | - | 1 |
| | | Paris | | |
| < 300 | 9 | - | 2 | 2 |
| 300-500 | 2 | 2 | 3 | - |
| > 500 | 2 | - | 1 | - |
| **Asuka 12236** | | | | |
| < 300 | 4 | 4 | 4 | 2 |
| 300-500 | 2 | - | - | 3 |
| > 500 | 1 | - | - | 1 |
| **Total** | | | | |
| < 300 | 21 | 7 | 7 | 4 |
| 300-500 | 7 | 4 | 3 | 3 |
| > 500 | 4 | 3 | 1 | 2 |
| **sum** | **32** | **14** | **11** | **9** |

***Table 1:*** *The 66 chondrules with fine-grained rims analyzed in this study divided according to CM chondrite section, three-dimensional size and latitude within the chondrules (i.e., polar, midlatitude or equatorial). Nine of these chondrule planes could not be defined. Note that most of these chondrules have irregular shapes and are not perfectly spherical (see main text for details). As such, these diameters should be taken as approximations.*

*2.2 Scanning electron microscopy*

Polished thin sections of Paris, Asuka 12236, and Maribo were examined using a Zeiss EVO 15 scanning electron microscope (SEM) equipped with a Bruker X-Trace energy-dispersive X-ray spectroscopy (EDS) detector at the Centre for Star and Planet Formation, University of Copenhagen. The samples were analyzed uncoated in variable-pressure mode. Backscattered electron (BSE) images were acquired both at section-wide scale and at higher magnification for individual chondrules and their surrounding rims. Chondrules spanning a range of sizes and morphologies were selected to ensure representative sampling (**Table 1**). High-resolution BSE imaging and Mg-Fe-Si elemental maps of selected inclusions and rims were performed with an accelerating voltage of 20 kV and beam currents between 1.0 and 5.0 nA, with a constant working distance of 9.5 mm (see supplementary figures S1-S9).

*2.3 X-ray Computed Tomography (XCT)*

X-ray micro–computed tomography (μCT) using a Zeiss Xradia 520 Versa system at the Technical University of Denmark (DTU) was performed on each section at low and high resolution (see ***supplementary videos***). For the latter, a total of 15 volumes (each approximately 3 × 3 × 3 mm) were imaged: seven volumes were scanned from Maribo, three from Asuka 12236, and five from Paris. The positions of these scans are shown in the overview BSE maps of each section (***Fig. S3-6***). Scans were conducted using accelerating voltages of 80 kV for Maribo and Asuka 12236, and 120 kV for Paris, reflecting differences in sample density and X-ray attenuation. Beam hardening was mitigated through the use of appropriate filters depending on the modal abundances of dense phases such as metal: an LE2 filter for Maribo, an LE4 filter for Asuka 12236, and an HE2 filter for Paris. All high resolution datasets were acquired at an isotropic voxel size of 3 μm. Exposure times were adjusted to optimize signal-to-noise ratios and ranged from 4 s for Maribo, 5 s for Asuka 12236, and 10 s for Paris. Tomographic reconstructions were performed using the instrument's standard reconstruction pipeline, yielding three-dimensional volumes suitable for quantitative analysis of internal textures, including the distribution and morphology of chondrules, matrix, and dense phases such as metals and sulfides. Data reduction of these volumes was carried out using Fiji (ImageJ) and Dragonfly, which we used for the 3D construction of chondrules and their rims, as well as to determine the sectioning plane of each chondrule and its size. During this determination the assumption was made that each chondrule is approximately spherical. The sectioning planes for each chondrule were divided into polar, midlatitude and equatorial, where the latter reflects the center plane of a chondrule, midlatitude represents planes between 25 % and 50 % (center) radius and polar reflects 0-25 % radius.

*2.4 fs-LA-ICP-MS*

Major and trace element concentrations of 66 chondrules and their surrounding FGRs were measured by femtosecond laser ablation inductively coupled plasma mass spectrometry (fs-LA-ICP-MS) using an ESI NWRFemto laser ablation system coupled to a Thermo Scientific iCAP RQ ICP-MS at the Centre for Star and Planet Formation, University of Copenhagen and using a 257 nm wavelength. Prior to each analytical session, the laser ablation chamber was purged with $N_2$ at a flow rate of 1500 mL $min^{-1}$ for approximately 15 minutes to remove atmospheric gases. The laser was subsequently warmed for a minimum of 10 minutes to ensure stable output energy.

Before data acquisition, the ICP-MS was tuned using the NIST SRM 612 glass standard to optimize signal sensitivity and stability. Gas flow rates were adjusted to minimize oxide formation, monitored via the percentage of $^{232}Th^{16}O^+$ from the total $^{232}Th^+$ counts, which was typically maintained at <0.4 %. During tuning, the $^{232}Th/^{238}U$ ratio was kept at 1.00 ± 0.03. Helium was used as the carrier gas, with a flow rate of 800 mL $min^{-1}$. Laser ablation was performed at a fluence of 6–7 J $cm^{-2}$, using a 20 μm spot size, a repetition rate of 100 Hz, and a scan speed of 40 μm $s^{-1}$. Elemental distribution maps were acquired in line-scan mode, with a 50% overlap between adjacent lines (10 μm) to enhance spatial resolution and minimize undersampling effects. The following isotopes were measured during all analytical sessions: $^{23}Na$, $^{24}Mg$, $^{27}Al$, $^{29}Si$, $^{31}P$, $^{39}K$, $^{44}Ca$, $^{49}Ti$, $^{51}V$, $^{52}Cr$, $^{55}Mn$, $^{57}Fe$, $^{59}Co$, $^{60}Ni$, $^{63}Cu$, $^{66}Zn$, $^{71}Ga$, $^{73}Ge$, $^{77}Se$, $^{85}Rb$, $^{90}Zr$, $^{95}Mo$, $^{107}Ag$, $^{111}Cd$, $^{118}Sn$, $^{125}Te$, $^{182}W$, $^{197}Au$, $^{208}Pb$, where the total dwell time was 350 ms. The analyses were started after the ICP-MS plasma was turned on >15 minutes to stabilize the background signal.

Analyses were conducted using a sample–standard bracketing approach, with standards measured before and after each sample block, where for the latter, we maintained a maximum analysis time of two hours to ensure representative standard correction. This means that we started with a standard block analysis, followed by a sample block and the final sample block followed by a standard block. A typical sample block would consist of 2-3 chondrules. Typically, we experienced the most significant loss of drift between the first two standard blocks, after which the signal intensity stabilized. This drift was accounted for using an autospline function in Iolite v.4. Reference materials included NIST SRM 610, 612 and BIR-1G glasses, as well as a pressed powder pellet of the geological reference material BHVO-2.

Data reduction was performed offline using Iolite v.4 and the 3D trace element data reduction scheme (Paton et al., 2011; Paul et al., 2012, 2023). The BIR-1G standard was used to normalize standard yields, which computes a median correction factor from elements shared between each calibrant and a primary reference material, then applies this single factor to all elements for that calibrant to compensate for relative ablation yield (matrix) differences. All standards were internally normalized to their Fe contents. Because Fe (and other element) concentrations vary significantly among chondrule phases, an alternative internal normalization strategy was applied to the line scan images. For each distinguishable phase (i.e., core, sulfide, metal and rims), a phase-specific internal standard criterion was defined based on the signal intensity of characteristic elements. For example, Zn and Ge were used to distinguish core from rim and Ni was used to recognize metal. Elemental concentrations within regions meeting a given phase criterion were normalized such that the sum of measured elements equaled 100 wt.oxide% (and wt.% for metal and sulfide), thereby minimizing matrix effects and improving inter-phase comparability (Paul et al., 2023).

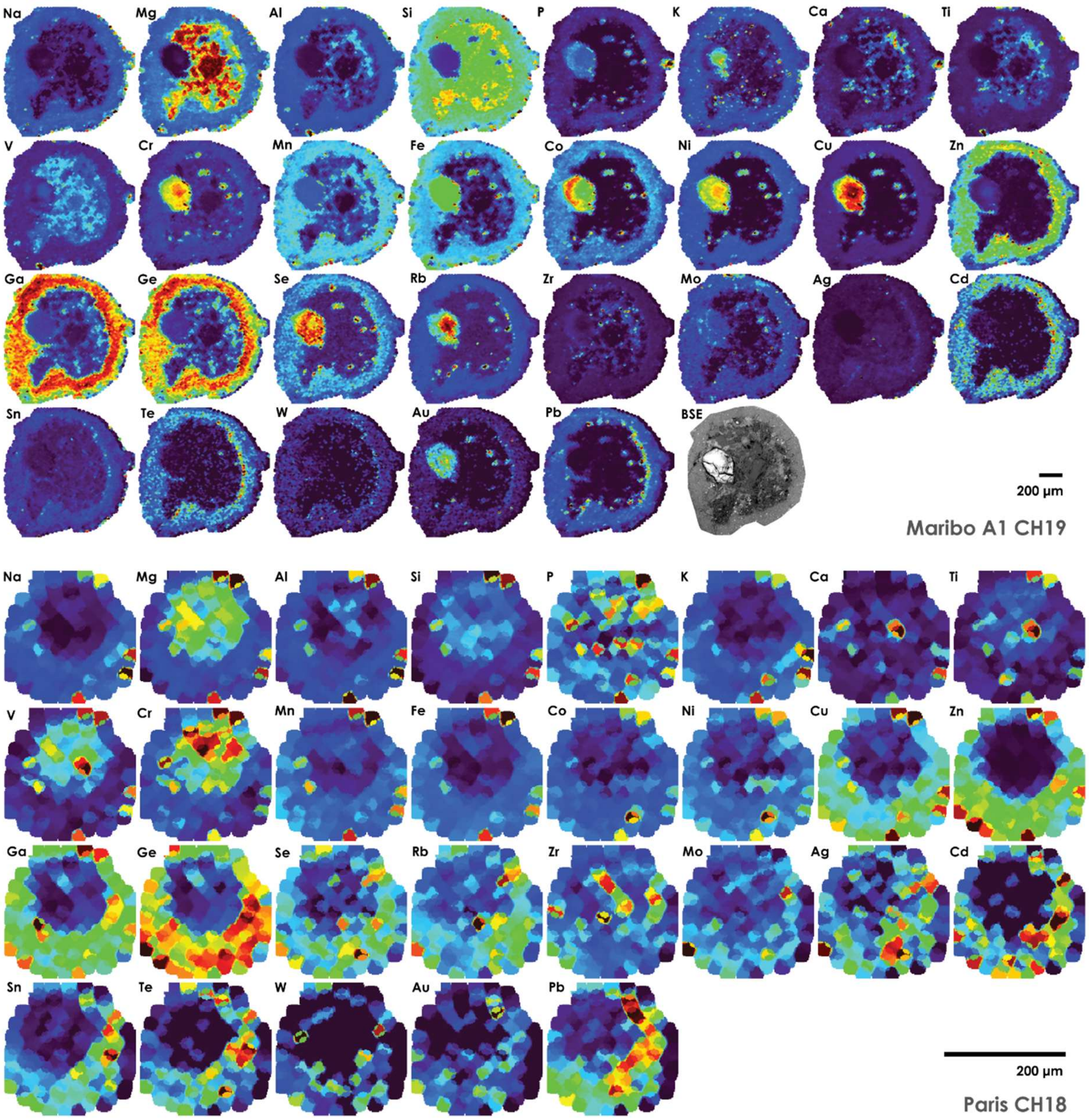

***Figure 2:*** *Elemental heat maps from fs-LA-ICP-MS using a relatively large chondrule (>500 µm, Maribo A1 CH19) and a small chondrule (~100 µm, Paris CH18), both with fine-grained rims, as examples, in which elements are most concentrated in red hues and most depleted in blues. Note the clear distinctions between chondrules and FGRs, especially for Zn, which is used as criterium in sampling chondrules versus rims in our dataset. Where the large chondrules allow us to see elemental variations between mineral phases, the spatial resolution (i.e., 20 µm spot size) is too low to observe this in very small chondrules.*

A key advantage of the imaging mode employed here is its ability to resolve the spatial distribution of elements among chondrule phases—such as sulfides, metal, olivine, and mesostasis—as well as chemical zonation within individual chondrules. In contrast to laser spot analyses, this approach yields more representative bulk compositions, at least for the two-dimensional section of the analyzed chondrule. Moreover, unlike electron microprobe or SEM analyses, LA-ICP-MS provides quantitative trace-element concentrations at (sub)-ppm levels. This gain in chemical sensitivity comes at the expense of spatial resolution, which is limited to approximately the laser spot size (~20 µm), whereas SEM allows semi-quantitative major-element analyses at the micrometer scale.

These trade-offs are illustrated in **Figure 2**, which shows LA-ICP-MS elemental maps of both relatively large chondrules (>500 µm) and the smallest chondrules analyzed here (~100 µm). The larger chondrules preserve resolvable internal features, including individual olivine crystals, mesostasis, and sulfides, as well as chemical zonation within sulfide grains (e.g., Co maps) and core-to-rim variations at the chondrule scale (e.g., Si maps). The smaller chondrules show limited internal chemical heterogeneity at the scale resolved by LA-ICP-MS but are nevertheless included in the bulk compositional analyses of chondrules and their rims together with the larger chondrules.

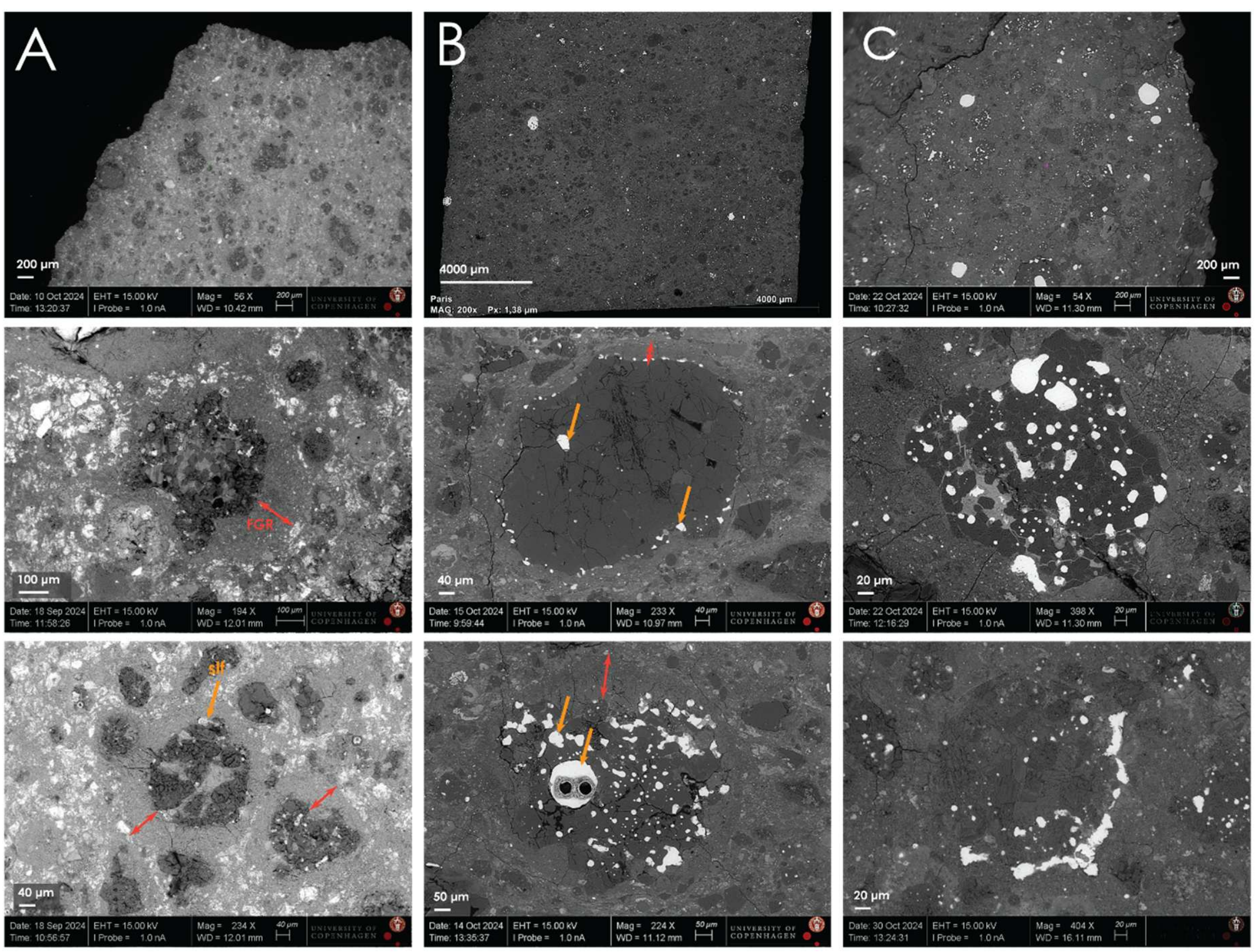

***Figure 3:*** *Back scattered electron images of column A: Maribo, column B: Paris and column C: Asuka 12336. Top panels show large scale texture of the CM chondrites. Red arrows mark the width of the FGRs, orange arrows point to sulfides, whereas yellow arrows point to examples of metal grains (white in BSE). Note the lack of metal grains in Maribo, their thick FGRs compared to Paris and Asuka 12236, as well as the irregular shapes of the chondrules that resemble 'grape-bunch' textures (Müller et al., 1969). The two pits in the bottom B panel are laser ablation spots from* (van Kooten et al., 2022).

## 3. Results

*3.1 Petrographic description from SEM*

In addition to the detailed petrographic descriptions provided by previous studies (Haack et al., 2012; Hewins et al., 2014; Kimura et al., 2020), we focus here on observations that distinguish the investigated samples from one another and from typical CM chondrite characteristics. A key difference among the samples concerns their metal content. Whereas Paris and Asuka 12236 contain abundant FeNi metal grains within chondrules and armored metal rims surrounding them, Maribo lacks observable metal at this scale and contains only rare sulfides within chondrules (**Fig. 3**).

FGRs surrounding Maribo chondrules are generally thicker (approximately 100 µm around ~300 µm chondrules) than those observed in Paris and Asuka 12236, where rim thicknesses typically do not exceed ~50 µm; Asuka 12236 exhibits the thinnest rims overall. Despite these differences, chondrules in all three CM chondrites commonly display irregular morphologies, with well-rounded chondrules being the exception rather than the rule. In most cases, however, FGRs infill chondrule surface irregularities and form relatively smooth, rounded composite objects.
Another general characteristic of type I porphyritic chondrules in all samples is their dense population of forsterite grains, with only minor amounts of mesostasis occupying the interstitial regions. As noted in previous studies of Paris and Asuka 12236 chondrules (Kimura et al., 2020), a significant fraction of the chondrules in the sections contains altered mesostasis and oxidized metal grains (**Fig. 3** and **Fig. S1-6**). Notably, the fine-grained rims appear unaltered, whereas the encapsulated chondrules show evidence of oxidation of the metal and leaching of Fe into the mesostasis, resulting in a lightening of this phase in BSE imaging mode. Note that we exclude type II (FeO-rich) chondrules from this study and focus on the type I chondrules, mainly because we did not find enough type II chondrules in these sections to provide a statistically relevant dataset at this point. All BSE images of the chondrules in this study are provided in the Supplementary Materials (**Fig. S1-6**).

*3.2 Average 2D chondrule and rim compositions of CM chondrites*
We report compositions of individual bulk chondrules and FGRs (***Supplementary Table R1***), as well as averaged chondrule and rim compositions for Maribo, Paris, and Asuka 12236 (**Table 2**). **Figure 4** presents the average FGR and bulk chondrule compositions of Maribo (Sections A and B), Paris, and Asuka 12236, normalized to CI chondrites (Lodders, 2003) and plotted according to their relative volatility (Wood et al., 2019).

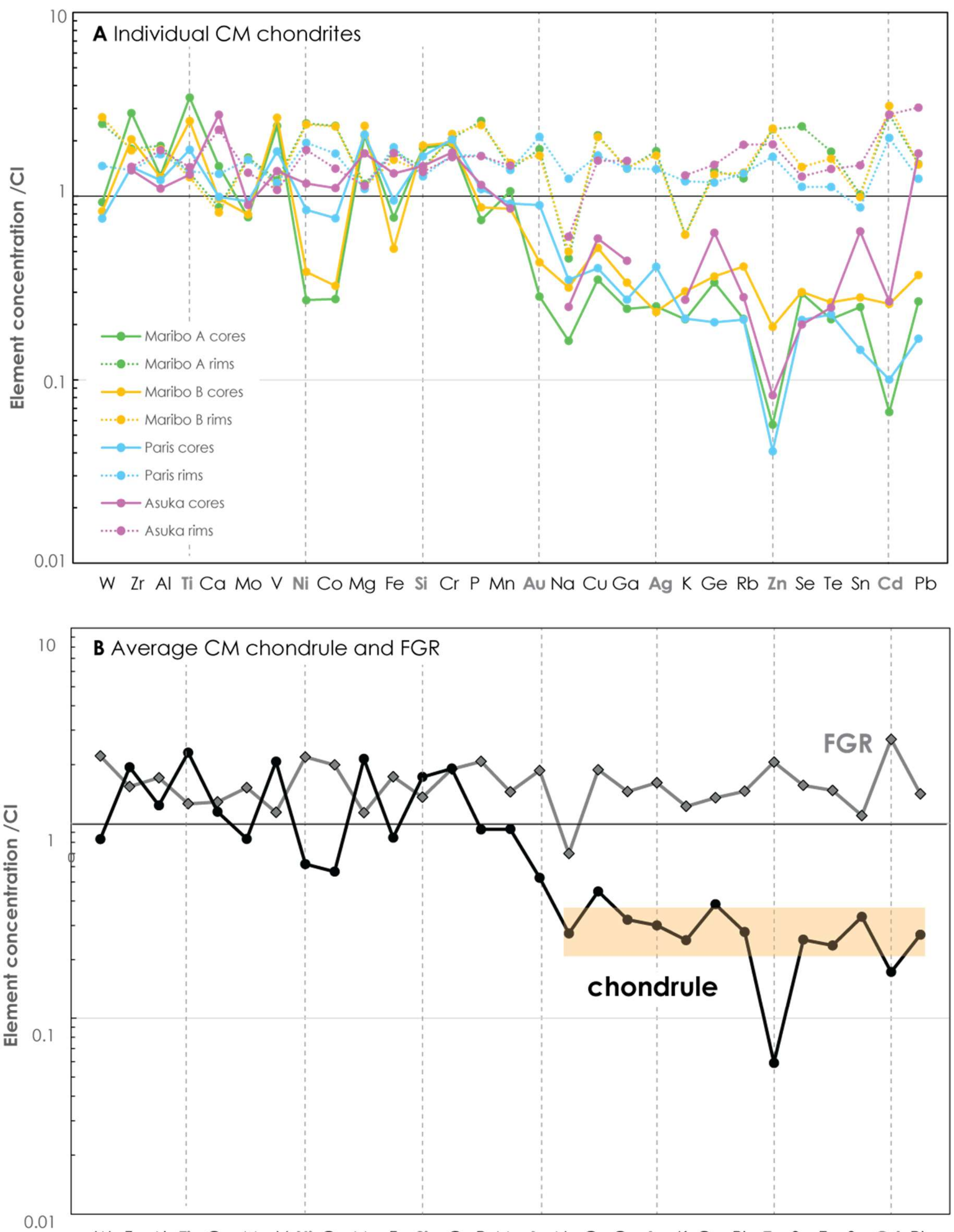

***Figure 4: A**) The averaged chondrule (cores) and FGRs (rims) compositions for each CM chondrite sample, with element concentrations normalized to CI chondrites* (Lodders, 2003) *and plotted according to their relative volatility* (Wood et al., 2019)*, where more volatile elements plot to the right. Gold contents (Au) for Asuka 12236 have not been plotted, since these values are contaminated by gold coating of the sample. **B**) The average chondrule and rim composition for all chondrules in this study (n=66), with FGRs at 1.6 × CI chondrite and an MVE plateau at 0.29× CI (orange bar including 2σ error). Note that Zn and Cd are excluded from this plateau.*

Rim element concentrations are slightly elevated but highly uniform relative to CI chondrites among the three CM chondrites, particularly for Paris (1.5 ± 0.3 × CI; 1σ), with comparable averages for Maribo (1.6 ± 0.6 × CI) and Asuka 12236 (1.6 ± 0.5 × CI). This elevation relative to CI chondrites reflects the sum normalization of the rim composition in the absence of water (i.e., FGRs essentially equal water-free CI chondrites). Maribo FGRs show notable depletions in Na and K, previously attributed to terrestrial weathering (Haack et al., 2012), and are relatively enriched in siderophile and chalcophile elements (e.g., W, Ni, Co, Cr, Zn and P) compared to Paris and Asuka 12236 (**Table 2**). The 2σ variations of the CI-normalized elemental concentrations are typically <0.2 for the average FGRs (see also **Table 2**). Both Paris and Asuka 12236 FGR compositions generally agree with 'matrix' compositions previously defined using LA-ICP-MS and electron microprobe, respectively (Hewins et al., 2014; Patzer et al., 2023).

Average bulk chondrule compositions from Asuka 12236, Paris, and Maribo are remarkably homogeneous within 2σ uncertainties (**Fig. 4A, Table 2**). The most pronounced differences occur among siderophile elements, with Maribo chondrules exhibiting stronger depletions in Ni, Co, and Fe relative to Paris, and especially relative to Asuka 12236, which approaches CI abundances. Both chondrules and rims of Asuka 12236 have notably higher concentrations of Ca and Pb compared to Maribo and Paris. This is perhaps related to calcite precipitation in CM chondrites during terrestrial weathering, although Asuka 12236 has weathering grade A. Major and minor element concentrations of Paris and Asuka 12236 have previously been determined using electron microprobe (Patzer et al., 2023) and generally agree well with data obtained in this study (**Fig. 5A**). Siderophile element abundances are typically higher for chondrules from Asuka 12236 and Paris, which is why the CM chondrule average from Patzer et al. (2023) – that does not include siderophile element depleted Maribo chondrules – falls above the 1:1 ratio line in **Figure 5A**. In addition, Ti concentrations in Maribo chondrules are generally higher than in the other CM chondrules.

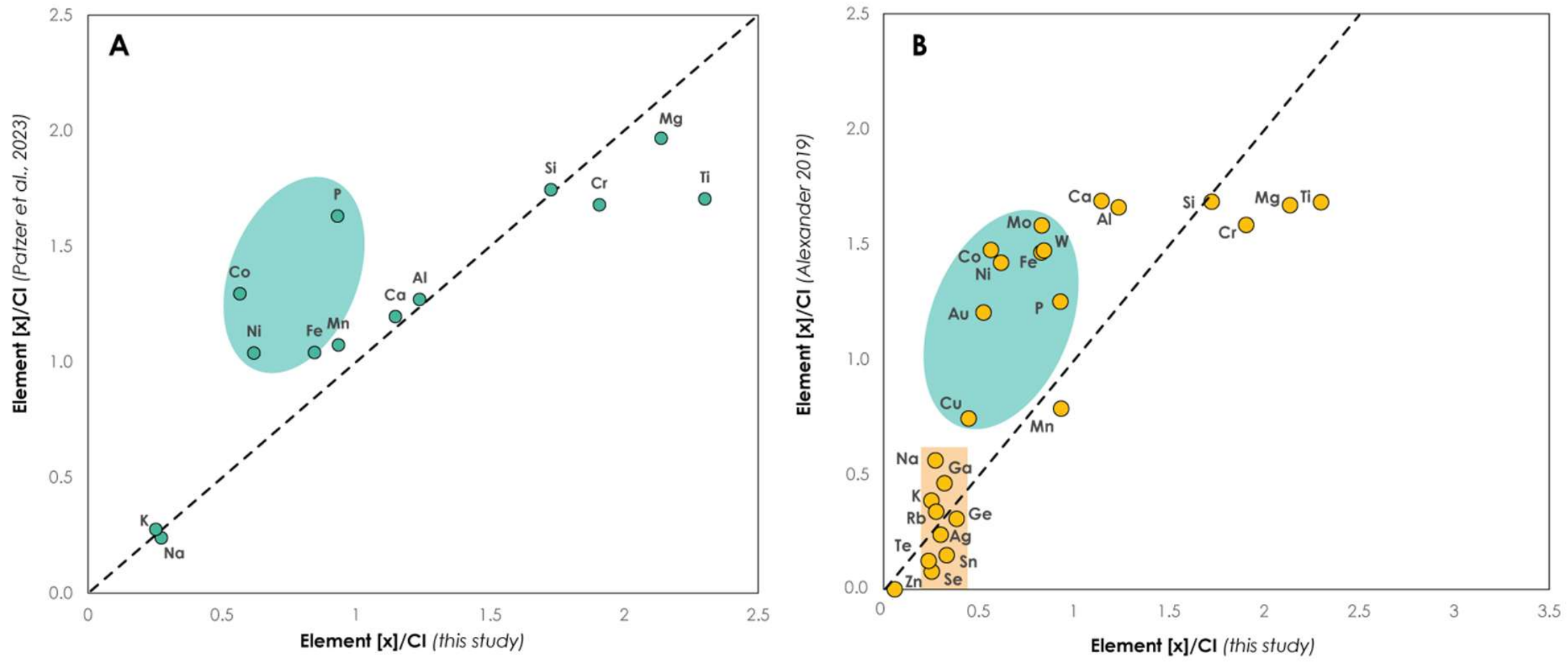

*__Figure 5: A__) Elemental abundances normalized to CI chondrites for averaged CM chondrules from this study (see also Fig. 4B and Table 2) plotted against the average chondrule composition obtained by Patzer et al. (2023). The dashed line represents the 1:1 ratio. __B__) The same as panel A but plotted against modelled chondrule composition from Alexander (2019). The green ellipse reflects siderophile elements and the orange rectangle represents the MVE plateau.*

Overall, refractory lithophile elements (T > 1300 K) in the chondrules are enriched to levels similar to those of the matrix (~1.6 × CI), whereas moderately volatile elements (1300–1100 K) approach solar abundances and decline to a plateau at 0.29±0.05 × CI at 50% condensation temperatures <1000 K. Within this plateau, chondrules show a significant depletion in Zn (0.06 × CI) and to lesser extent in Cd (0.17 × CI). This amounts to 18.3±7.5 µg/g and 0.12±0.09 µg/g of Zn and Cd in CM chondrules, respectively (**Table 2**). This low Zn concentration in chondrules agrees well with model predictions based on the bulk Zn concentrations and chondrule to matrix ratios of carbonaceous chondrites (Alexander, 2019; Zhang and Grewal, 2026), where the Zn content of the non-matrix fraction (i.e., chondrules) is near-zero. However, there are some important differences between the predicted chondrule composition from Alexander (2019) and our analysis (**Fig. 5B**). First, there is no predicted depletion of siderophile elements relative to CI chondrites in the modelled chondrule composition, whereas Maribo chondrules, and Paris to a certain extent, show marked siderophile element depletions (**Fig. 4A**). Second, the MVEs in CM chondrules from this study follow a plateau, rather than a continuous depletion with increasing volatility as is produced in modelled chondrule compositions (Alexander, 2019; Zhang and Grewal, 2026). As such, MVEs with relatively low volatility are overproduced in the model chondrules, whereas MVEs with higher volatility are underproduced (**Fig. 5B**). Finally, major elements such as Ca, Al and Mg have poor fits with observed chondrule compositions in this study, where Ca and Al are overestimated in the model and Mg underestimated.

*3.3 Assessing the potential size and sectioning bias and secondary alteration*

We have analyzed chondrule compositions within the main size range observed in CM chondrites (~100 to 900 µm) and at different chondrule planes (i.e., polar, midlatitude and equatorial, **Table 1**). This allows us to assess potential compositional size and sectioning biases in, for example, sampling large over small chondrules or sampling chondrules at preferential planes exposed during sectioning. In **Figure 6A** we have plotted the chemical compositions of chondrules according to their respective size bins, with elements plotted according to their relative volatility, similar to **Figure 4**. These chondrule sizes reflect the diameter of the chondrule taken from the laser ablation data using Zn as a criterium to distinguish the chondrule from the FGR. As such, these sizes may not reflect the actual size of the chondrule, but the sectioned size. Hence, in **Figure 6B**, we also plot the chondrules according to their 3D size bins, as taken from the high resolution XCT scans. These sizes are calculated as averaged X,Y,Z diameters, since CM chondrules can contain highly irregular shapes. For elements with $T_{50}$ > 1000 K, no significant variations are observed related to the 2D size of the chondrules (**Fig. 6A**). At the MVE plateau, the larger chondrules appear systematically more depleted than the smaller chondrules. When using the 3D sizes of chondrules, this observation does not disappear, but the largest offset is between more MVE-depleted chondrules >300 µm and relatively enriched smaller chondrules with diameters <300 µm (**Fig. 6B**). The actual sizes of chondrules also reflect metal-depleted larger chondrules, versus metal-rich smaller chondrules. This may be related to the fact that Maribo chondrules are more depleted in siderophile elements and also contain some of the largest chondrules in our sample selections. Finally, in **Figure 6C** we plot the chemical compositions of chondrules according to their sectioned plane. We find no significant differences between polar, midlatitudinal or equatorial planes of chondrules.

*3.4 Magnesium/silicon ratios of chondrules and their rims*

We investigated Mg–Si systematics in CM chondrules and their fine-grained rims (FGRs) (**Fig. 7**), which are central to the chondrule–matrix complementarity debate (e.g., Hezel and Palme, 2010; Ebel et al., 2016; van Kooten et al., 2024). This framework highlights the contrast between sub-chondritic Mg/Si ratios in carbonaceous chondrite matrix and super-chondritic ratios in chondrules (Hezel and Palme, 2010). We report Mg and Si concentrations for 66 chondrules and their FGRs, shown in **Figure 7** together with the CI chondrite composition and Mg/Si ratio line.

***Figure 6**: CM chondrule compositions according to their 2D and 3D size and chondrule plane, with element concentrations normalized to CI chondrites* (Lodders, 2003) *and plotted according to their relative volatility* (Wood et al., 2019). ***A**) Chondrule compositions according to their 2D size in micrometer, **B**) according to their 3D size and **C**) according to their chondrule plane.*

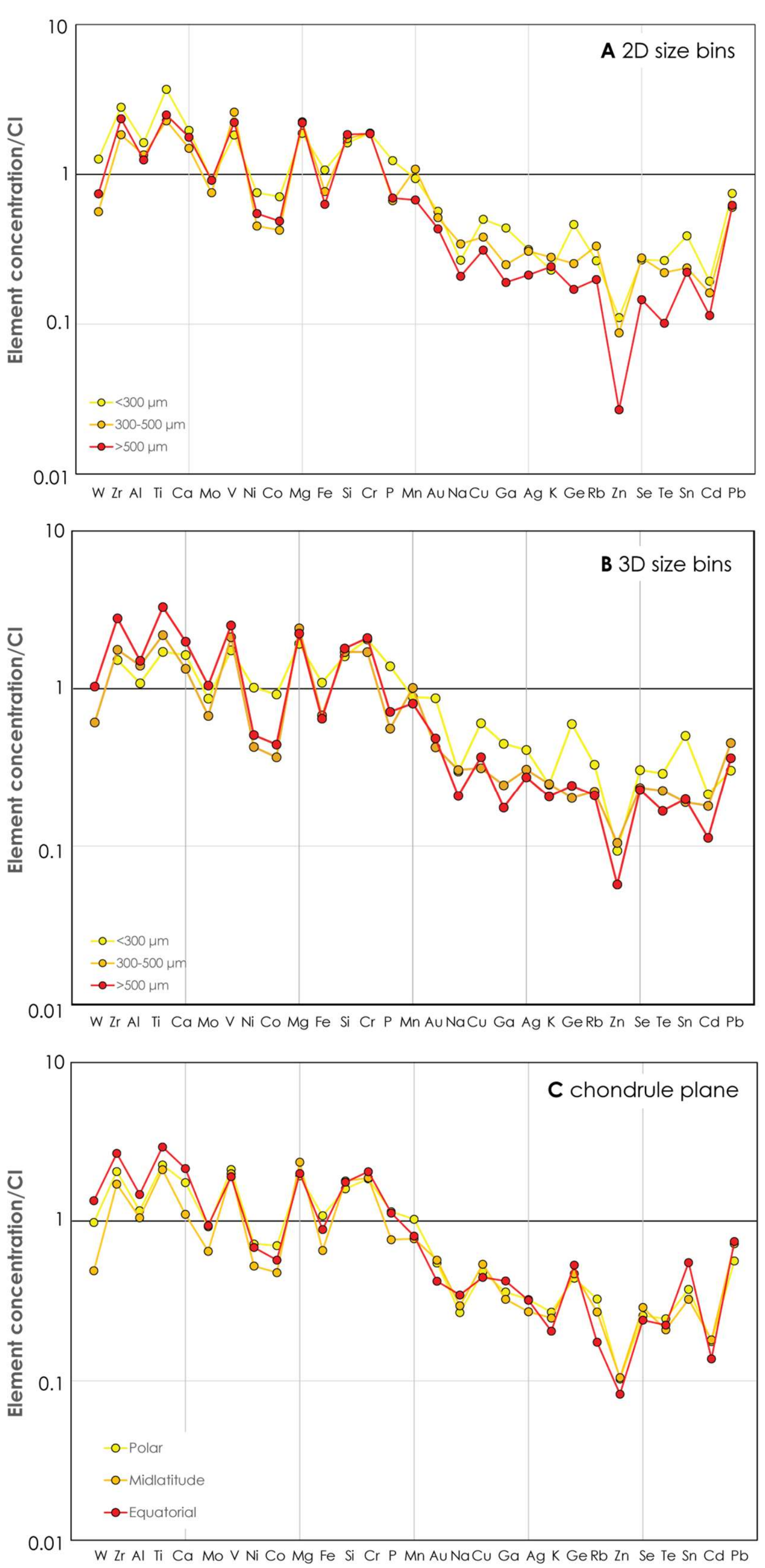

The FGRs define a tight cluster in Mg–Si space (**Fig. 7A**), with average concentrations of 10.9 ± 0.05 wt.% Mg and 14.5 ± 0.1 wt.% Si. These values are clearly distinct from CI chondrite compositions (Mg = 9.6 ± 0.01 wt.%; Si = 10.7 ± 0.04 wt.%; Lodders et al., 2003), confirming that CM chondrite FGRs—and by extension the matrix—exhibit sub-chondritic Mg/Si ratios (Hezel and Palme, 2010; Ebel et al., 2016). In contrast, bulk chondrule compositions show significantly greater variability, with average concentrations of 19.9 ± 0.6 wt.% Mg and 18.1 ± 0.4 wt.% Si, corresponding to nearly an order of magnitude more scatter than observed in the surrounding rims.

Individual chondrule compositions lie along the CI chondrite correlation and extend toward super-chondritic Mg/Si ratios approaching the composition of forsterite (**Fig. 7A**). Highly metal-rich chondrules and individual chondrule metal grains from Paris and Asuka 12236 were also analyzed; their Mg–Si compositions plot along the CI correlation line.

## 4. Discussion

### *4.1 Is our dataset representative of the CM chondrule composition?*

A critical consideration when interpreting our CM chondrule dataset is whether it accurately reflects the broader CM chondrule population. Our study encompasses 66 bulk chondrules analyzed via laser ablation line scans, providing major, minor, and trace element compositions across a wide volatility range. To our knowledge, this constitutes the most extensive trace element dataset for CM chondrules to date. Previous investigations have largely relied on SEM or EMPA, which allow for faster analyses of a larger number of chondrules but are typically limited to major and selected minor elements (Ebel et al., 2016). While such techniques have proven very valuable for assessing bulk major element trends, they do not capture the full spectrum of moderately volatile and trace elements, including the MVE plateau observed here, which is essential for comparisons with modeled chondrule compositions (Alexander, 2019; Patzer et al., 2023).

#### *4.1.1. Chondrule size and sectioning bias*

In our dataset, we observe limited variability between chondrules from different CM meteorites (**Fig. 3**) and no systematic differences associated with chondrule size or apparent sectioning plane (i.e., polar versus equatorial). This suggests that our sampling captures the overall compositional range of CM chondrules. Nevertheless, one notable exception is the siderophile element depletion observed in Maribo chondrules relative to Paris and Asuka 12236. For this reason, larger chondrules in **Figure 6B** are seemingly more siderophile element depleted. However, since we typically sampled larger chondrules from Maribo (>500 µm, **Table 1**), this size bias effect is artificial.

Notably, the overall chemical composition of CM chondrules from this study shows limited evidence of the previously described sectioning bias (**Fig. 6C**), where polar type I porphyritic chondrules exhibit a different mineralogy (i.e., low-Ca pyroxene dominated) relative to their equatorial planes (forsterite dominated; Friend et al., 2016; Barosch et al., 2020). Friend et al. (2016) have observed low-Ca pyroxene zonation around 80 % of the CM chondrules, including 11 chondrules from El-Quss Abu Said and 30 chondrules from Jbilet Winselwan. In this study, we find significantly lower abundances of chemically zoned chondrules from Mg-Si elemental maps: 4 out of 42 in Maribo, 5 out of 15 in Paris and 2 out of 13 in Asuka 12236 (**see supplementary** **figures 7-9**). As such, the paucity of zonation (averaging ~16 % of zoned type I chondrules) found in these elemental maps agrees with that of the LA-ICP-MS data and the lack of sectioning bias observed between polar, midlatitudinal and equatorial planes (**Fig. 6C**). A fraction of the chondrules in this study do exhibit potential zonation, but their low number and relatively thin zonation compared to the chondrule diameter is reflected in the averaged sectioned chondrule compositions. Moreover, against expectations, the polar chondrules in this study do not show a higher modal abundance of low-Ca pyroxene relative to the equatorial planes. The discrepancy between zoned CM chondrules observed by Friend et al. (2016) and in our study may reflect the degree of secondary alteration recorded in these samples. Although we have limited information on the alteration degree of El-Quss Abu Said (CM2 chondrite), Jbilet Winselwan is a heavily altered CM chondrite (CM2.1-2.4) that shows clear mineral zonation within the FGRs surrounding the zoned chondrules (Friend et al., 2016). Moreover, Jbilet Winselwan is a breccia of which some lithologies records thermal metamorphism up to 500 °C that have further altered its primary composition and mineralogy (Grady et al., 2014; Russell et al., 2014; Zolensky et al., 2016; King et al., 2019). As such, it is feasible that low-Ca pyroxene zonation in altered chondrites represents a secondary process, rather than a primary gas-melt interaction. We note that this may only pertain to observations related to CM chondrules and not to observations of ordinary chondrite chondrules with low degrees of secondary alteration (Barosch et al., 2019, 2020).

Critically, the paucity of low-Ca pyroxene zonation in unaltered CM chondrules does not negate the concept of gas-melt interaction during chondrule formation, since this is corroborated by the observation of epitaxial growth from zoned olivines in chondrules (Libourel and Portail, 2018), zoned isolated olivine grains (Jacquet et al., 2021) and limited mass-dependent isotope fractionation during evaporation (Marrocchi et al., 2018, 2024; Villeneuve et al., 2020). The fact that we do not see this gas-melt interaction dominantly reflecting as low-Ca pyroxene zonation in the CM chondrules may be because the chondrules are small enough in size to be entirely buffered by the gas.

Finally, consistent with our observations of low-Ca pyroxene rims in CM chondrules, we find only limited evidence for zonation of MVEs, particularly alkali elements, within the chondrules (**Supplementary Fig. 10**). Previous studies have reported enrichment of alkali elements at chondrule margins, which has been interpreted as the result of recondensation during gas–melt interaction (Jacquet et al., 2026). Similarly, mass-dependent Fe isotope systematics in CM chondrules have been attributed to evaporation and subsequent recondensation processes during chondrule formation (Hezel et al., 2018b).
An alternative explanation for alkali zonation invokes secondary alteration processes, whereby Na is mobilized from FGRs into the chondrules, which typically contain significantly lower Na concentrations – often by an order of magnitude (Grossman and Brearley, 2005; Alexander and Grossman, 2005; Jiang et al., 2021). Because the samples analyzed here are relatively unaltered CM chondrules, this could account for the absence of pronounced zonation in our observations. Another possibility is that the small size of CM chondrules allowed recondensation to affect the entire chondrule rather than being restricted to outer regions.
In CV chondrules, light isotope enrichments of MVEs have been interpreted as evidence for recondensation into the chondrule melt or onto the chondrule surface, followed by diffusion (Hellmann et al., 2020; Nie et al., 2021; Jacquet et al., 2026; Zhang and Grewal, 2026). Alternatively, these signatures may reflect the physical removal of mineral phases enriched in isotopically heavy MVEs (Pringle et al., 2017). However, distinguishing between these scenarios falls outside the scope of this study and remains challenging due to small sample sizes, low elemental abundances, and the redistribution of MVEs during secondary alteration. Future work could help resolve these ambiguities through isotopic analyses of MVEs – such as Zn, Rb, and K – in CM chondrules.

*4.1.2. Redistribution by secondary alteration*

Another important consideration concerns secondary alteration of the chondrules and their FGRs, and the potential for chemical modification during parent-body processing. Although we deliberately selected the least altered CM chondrites available, all CM samples exhibit some degree of aqueous alteration. Previous nanoscale investigations of FGRs in Paris, Maribo, and Asuka 12236 (Kimura et al., 2020; Leroux et al., 2015; van Kooten et al., 2018) documented only limited evidence for aqueous overprinting (see section 2.1.1: Asuka 12236). This relative preservation forms the basis for their low petrologic subtype assignments within the CM group.

In contrast, a substantial fraction of chondrules—even in Asuka 12236 (~50%)—shows evidence of alteration, including oxidation of primary metal grains and the development of phyllosilicates (Kimura et al., 2020). Such modification raises the possibility that primary chemical signatures in CM chondrules and their FGRs may have been disturbed by secondary exchange processes.

However, we observe no systematic evidence for chemical exchange between chondrules and their surrounding rims or matrix. Specifically, the compositions of the least altered chondrules in Asuka 12236 are indistinguishable from those of more altered chondrules in the same meteorite, as well as from chondrules in Paris and Maribo (**Fig. 4A**). If aqueous alteration had facilitated significant element mobility between chondrules and FGRs, compositional offsets would be expected between these populations.

Two observations argue against such secondary chondrule–matrix exchange. First, our distinction between chondrule cores and FGRs relies in part on Zn, an element known to be readily mobilized during aqueous alteration. Any Zn enrichment of outer chondrule domains via influx from the FGR would shift those regions beyond our compositional cutoff and thus exclude them from the defined chondrule core. Our analytical protocol therefore inherently filters out potentially exchanged boundary regions.

Second, although aqueous alteration affected many chondrules mineralogically, it appears not to have promoted measurable chemical exchange with the adjacent matrix. One plausible explanation is that alteration of some chondrules occurred prior to FGR accretion (**see section 4.4**), thereby limiting opportunities for subsequent rim–core interaction during fluid activity.

Collectively, these observations indicate that secondary alteration has not significantly modified the primary chondrule–rim chemical relationships captured in our dataset. We therefore conclude that the analyzed chondrule–FGR pairs are representative of the CM chondrule population and preserve primary compositional signatures.

*4.2 Resolving the Mg-Si chondrule matrix complementarity*

Metal-rich chondrules and metal grains analyzed at the spatial resolution of LA-ICP-MS define Mg and Si abundances that lie along the CI chondrite ratio line (**Fig. 7**), where the metal-rich chondrules reflect Mg and Si contents that were diluted by the high abundance of metal in these chondrules. This observation implies that removal of metal from initially CI-like chondrule precursors – most plausibly via metal expulsion during chondrule melting (Wasson and Rubin, 2010; Rubin, 2022; van Kooten et al., 2022)– drives bulk chondrule compositions toward higher Mg and Si abundances while preserving CI-like Mg/Si ratios. Such behavior is observed for a significant fraction of the chondrules (**Fig. 7**). The magnitude of this shift along the CI line correlates with the degree of siderophile element depletion (Ni, Co, Fe) in individual chondrules (**Fig. 4A**), with Maribo chondrules plotting farthest from CI composition and exhibiting the strongest siderophile depletions, consistent with extensive metal loss.

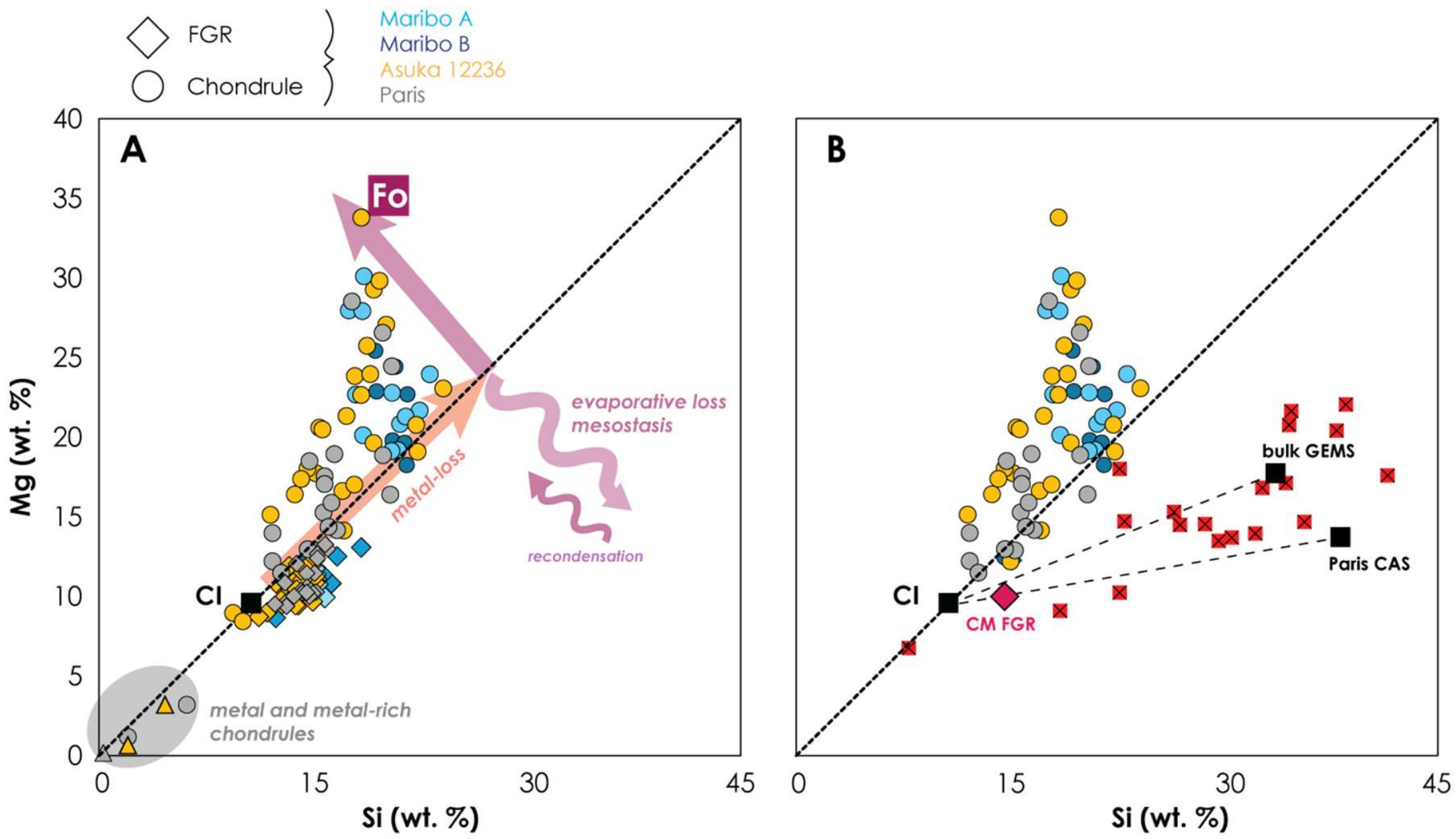

***Figure 7**: The Si versus Mg contents in weight percent of individual chondrules (circles) and FGRs (diamonds) from Maribo (blue), Paris (grey) and Asuka 12236 (yellow) plotted with the CI chondrite composition (black square) and the CI ratio line. **A**) Showing the process of Mg and Si chemical evolution during chondrule formation, assuming a CI chondrite starting composition and subsequent metal loss along the CI line and Si-dominated evaporation of mesostasis towards the forsterite composition (Fo). **B**) The mirroring sub-chondritic composition of bulk glassy embedded metals and sulfides (GEMS) and averaged chondritic anhydrous silicates (CAS) from Paris* (Leroux et al., 2015; Ohtaki et al., 2021; Schulz et al., 2024). *Also shown are individual GEMS compositions from CP-IDP U2-17B19* (Schulz et al., 2024), *where we recalculated atom percent elemental ratios to their respective weight percentages.*

At approximately ~2.5 × CI, this metal extraction trend reaches a limit, beyond which chondrule compositions deviate toward more Mg-rich values approaching the forsterite endmember (**Fig. 7A**). We interpret this deviation as reflecting evaporative loss of mesostasis, which is comparatively Si-rich, thereby driving the residual chondrule compositions away from the CI ratio line. This interpretation is supported by the observed correlation between moderately volatile element (MVE) depletion and increasing Mg/Si ratios of chondrules, quantified here using the V/MVE ratio (**Fig. 8**). This evaporative loss likely occurred at (near)-equilibrium conditions where the chondrules were buffered by the surrounding gas, resulting in partial recondensation of evaporated elements in and/or onto the chondrules (Libourel et al., 2006; Ebel et al., 2018; Jacquet et al., 2026). Collectively, these observations indicate that progressive changes in chondrule Mg/Si ratios reflect a generic chondrule formation pathway governed by a combination of physical metal loss and mesostasis evaporation. These processes may operate concurrently, producing the triangular distribution of chondrule compositions observed in Mg–Si space.

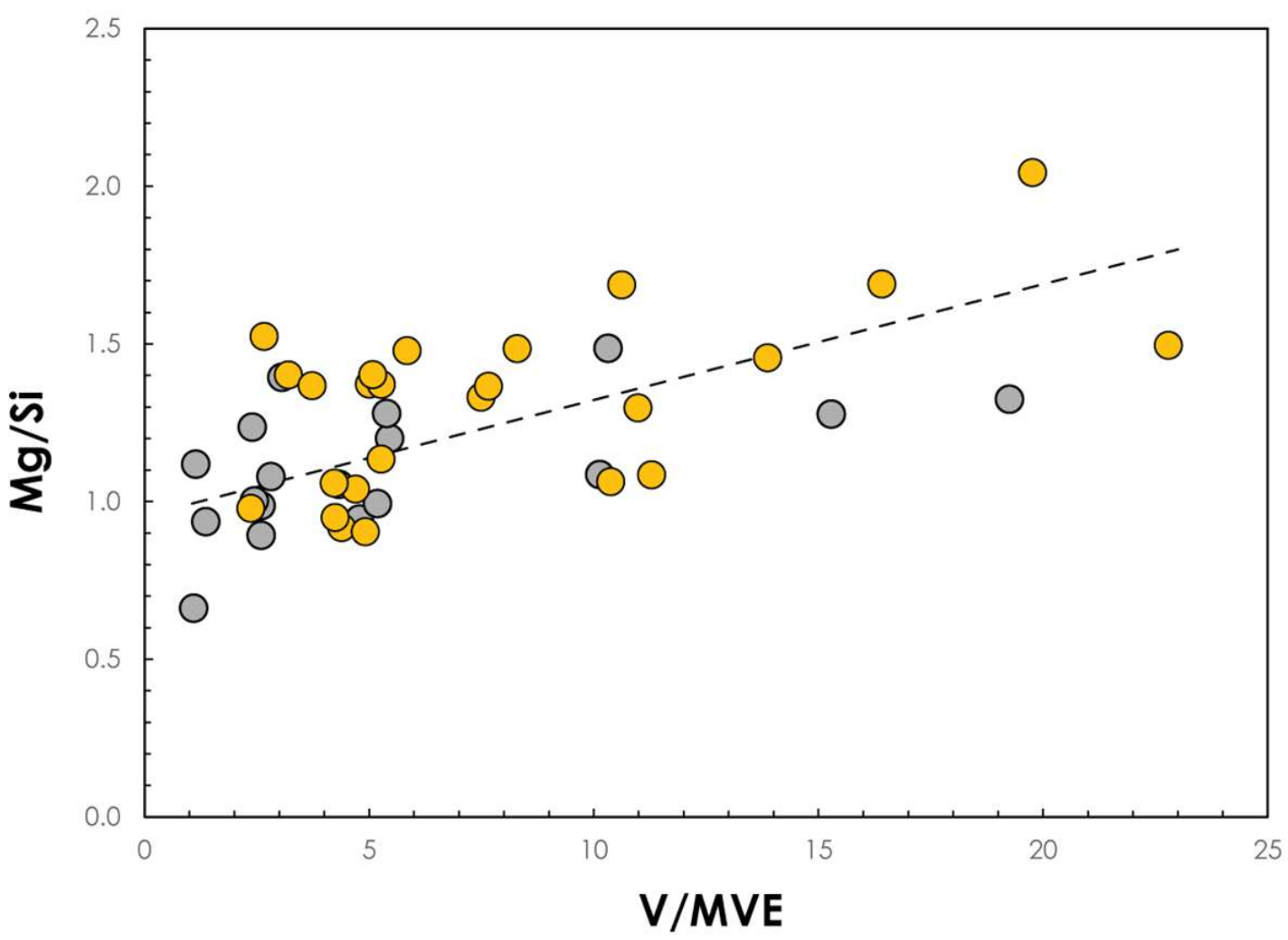

***Figure 8:*** *The volatile depletion of chondrules from Paris and Asuka 12236 expressed by V/MVE (averaged MVE plateau) versus the Mg/Si ratio. Maribo chondrules are excluded since their varying Na and K depletions from weathering artificially change the MVE plateau for individual chondrules. Note that this trend cannot resolve any recondensation of MVEs related to gas-melt interactions during chondrule formation, which has been previously observed in chondrules* (Jacquet et al., 2026)*, since this would move points downward on this trend.*

In contrast to the wide compositional range defined by chondrules, FGR compositions are tightly clustered in Mg–Si space, although they exhibit a tentative trend from CI composition toward slightly more Si-rich signatures (**Fig. 7A**). The average Mg/Si ratio of CM FGRs (0.76 ± 0.10, 1σ) is indistinguishable from that of chondritic porous interplanetary dust particles (Schramm et al., 1989; Ishii, 2019) and outer-disk dark clasts (Mg/Si = 0.67; (van Kooten et al., 2024b)), both commonly associated with cometary materials. This compositional similarity reinforces the interpretation that carbonaceous chondrites represent mixtures of cometary and CI-like precursor materials (van Kooten et al., 2024b).

Previous models addressing the genetic relationship between chondrules and matrix have interpreted chondrule–matrix complementarity in Mg–Si space as the consequence of forsterite redistribution during chondrule formation. In this view, evaporation of Mg-rich silicates from matrix precursors and subsequent incorporation into chondrules produced complementary Mg/Si signatures, although a quantitative physical mechanism capable of achieving this redistribution at scale has remained elusive (Hezel and Palme, 2010; Hezel et al., 2018a). Alternatively, it has been proposed that the observed complementarity instead reflects a more general feature of the chondrule-forming environment: the accretion of largely pristine, comet-like matrix material onto chondrules after their formation through a generic process that increases the Mg/Si ratio of chondrules. This interpretation is consistent with recent nucleosynthetic isotope constraints from outer Solar System dust reservoirs (van Kooten et al., 2024a).

Importantly, these two scenarios need not be mutually exclusive. They can be reconciled through consideration of the mineralogy and chemistry of GEMS (glassy embedded metals and sulfides) and GEMS-like materials. GEMS and related amorphous silicates are major constituents of chondritic porous interplanetary dust particles (CP-IDPs) and are widely regarded as key precursors to primitive chondrite matrices. Their compositions define a broad array in Mg–Si space that mirrors the compositional spread of chondrules. As illustrated in **Figure 7B**, published Mg and Si data for GEMS (Ishii, 2019; Ohtaki et al., 2021; Schulz et al., 2024) extend away from CI chondrite compositions and pass through the field defined by CM FGRs.
Chondritic amorphous silicates (CAS), preserved in the least altered matrices of meteorites such as Paris and Asuka 12236, are commonly embedded within GEMS-like materials (i.e., sulfide-bearing but metal-poor analogues of GEMS) and exhibit similar bulk compositions. The average Mg/Si ratio of CAS (n = 154; Ohtaki et al., 2021) is slightly lower than that of bulk GEMS (n = 193), and the mean CM FGR composition lies along a mixing line between CI-like material and CAS. Mass balance considerations indicate that CAS contributes ~14 wt.% to bulk CM FGRs, with the remaining ~86 wt.% represented by thermally unprocessed CI-like dust. Thus, CM fine-grained rims appear to record a hybrid origin, incorporating both pristine solar-composition material and a minor but significant fraction of chemically complementary condensate-like material. Our estimated abundance of chondrule-derived contributions to FGRs agrees well with Hellman et al., (2020), who estimated ~10 % of chondrule precursor dust added to CI-like matrix.
Experimental and petrographic studies have proposed that GEMS-like materials may form in high-temperature environments in the protoplanetary disk (e.g., Ishii et al., 2018; Keller and Messenger, 2011), and matrix has long been considered, at least in part, a by-product of chondrule formation (Ciesla et al., 2003; Desch et al., 2005; Huss et al., 2005). Supporting this possibility, amorphous silicates in pristine ordinary chondrite matrices are enriched in alkalis ($Na_2O$ ≈ 2 wt.%), Si, and S (Dobrică and Brearley, 2020)—elements preferentially lost during chondrule evaporation. Upon cooling, these volatile-rich vapors could recondense as fine-grained amorphous material and become reincorporated into the surrounding dust reservoir, thereby generating a matrix component chemically complementary to the newly formed chondrules. In this framework, the ~14 wt.% CAS-like component in CM FGRs may represent such a condensate fraction.

A similar argument may apply to comet-like materials. CP-IDPs and outer-disk dark clasts exhibit subchondritic Mg/Si ratios (Ohtaki et al., 2021; Schulz et al., 2024; van Kooten et al., 2024), consistent with the incorporation of GEMS-rich material. The origin of these GEMS—whether predominantly interstellar or formed in the solar nebula—remains debated (Bradley et al., 2022; Ishii, 2019; Keller and Messenger, 2011). The presence of chondrule-like objects in samples returned by Stardust demonstrates that high-temperature materials were transported to comet-forming regions, raising the possibility that at least some GEMS-like components in cometary matter represent by-products of chondrule-forming events. Conversely, GEMS in CP-IDPs and outer-disk clasts are closely associated with interstellar-derived organic matter and can be carbon-rich, features that may argue for a more complex or partially presolar origin. Resolving this issue will require detailed trace-element and isotopic characterization of GEMS and CAS across different reservoirs.

Taken together, our results support a model in which CM matrices are mixtures of (i) a condensate component chemically complementary to chondrules and (ii) thermally unprocessed, CI-like dust that retains a solar bulk composition and preserves presolar grains and organic matter. Such a hybrid origin provides a natural explanation for chondrule–matrix complementarity in Mg–Si space while maintaining the overall solar chemical and isotopic character of primitive chondritic matrices.

*4.3 The chondrule formation process: from small to big or big to small?*

Microchondrules are small silicate spherules (≲ tens of μm) that have been observed mainly within fine-grained rims and matrix of chondritic meteorites, and more recently in particles returned from asteroid Ryugu (Genge et al., 2025). Early work by Rubin et al., (1982) showed they can form by melting of dust aggregates, analogous to normal chondrules, indicating a similar fundamental origin.

Later studies on ordinary chondrites (Krot and Rubin, 1996; Krot et al., 1997) demonstrated that many microchondrules are closely associated with remelted chondrule margins, suggesting formation through localized secondary melting and dispersal of material from pre-existing chondrules. Consistent with this, (Bigolski et al., 2016) interpret ordinary chondrite microchondrules as byproducts of repeated, localized heating and recycling processes, formed either from chondrule surface melts or nearby dust aggregates in dynamic nebular environments. However, while relevant for ordinary chondrite chondrules, the textures and compositional features documented in this study challenge this interpretation in the context of CM chondrules and instead suggest an inverted formation sequence, in which microchondrules represent primary melt products that subsequently accreted and welded into larger composite chondrules.

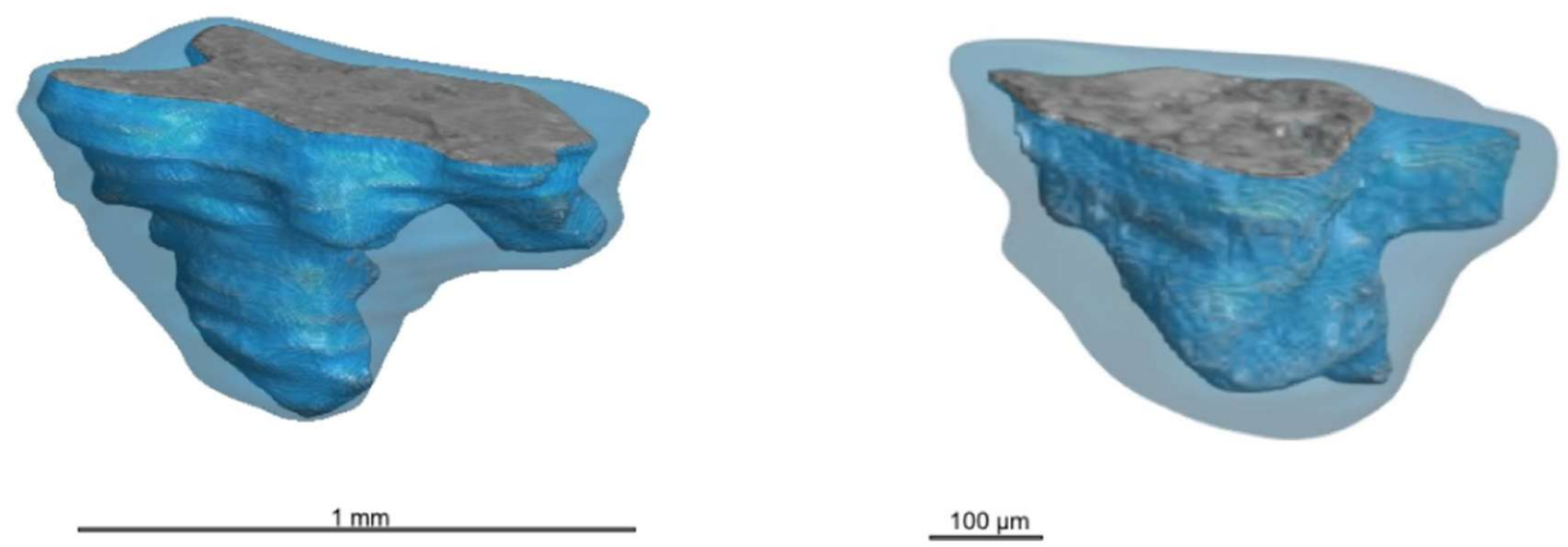

***Figure 9:*** *3D reconstructions of two grape-bunch chondrules from Maribo, where agglomerates of ~100 μm sized globules can be identified, in contrast to larger spherical objects. These grape-bunch textures are surrounded by the fine-grained rims (light blue) that fill the gaps between the microspherules. 3D videos of these objects are embedded in the supplementary materials.*

Rather than observing microchondrules preferentially attached to the surfaces of larger, texturally coherent chondrules, we find that many apparently singular large chondrules are internally composed of multiple, discrete microspherules. These internal structures are reminiscent of the "grape-bunch chondrules" originally described in C2 chondrites by Müller (1969). In our samples, this texture is evident not only in backscattered electron imaging (**Fig. 3 and Fig. S1-2**), but also in three-dimensional observations (**Fig. 9**) and in laser ablation trace-element maps (**Fig. 10**). The latter reveal subtle but systematic partitioning within individual chondrules, defined by Ni-, Co-, Cu-, Ga-, and Fe-enriched zones in Paris and Asuka 12236 that outline microspherule-scale domains (**Fig. 10**). In Maribo chondrules, we have identified inter-chondrule partitioning of microspherule domains separated by rim-like materials (**Fig. 11**). Such chemical heterogeneities are consistent with relict boundaries between individual microchondrules and may record the former presence of armored metal-rich rims in case of Paris and Asuka 12236 and fine-grained rims in case of Maribo, prior to coalescence (Wasson and Rubin, 2010).

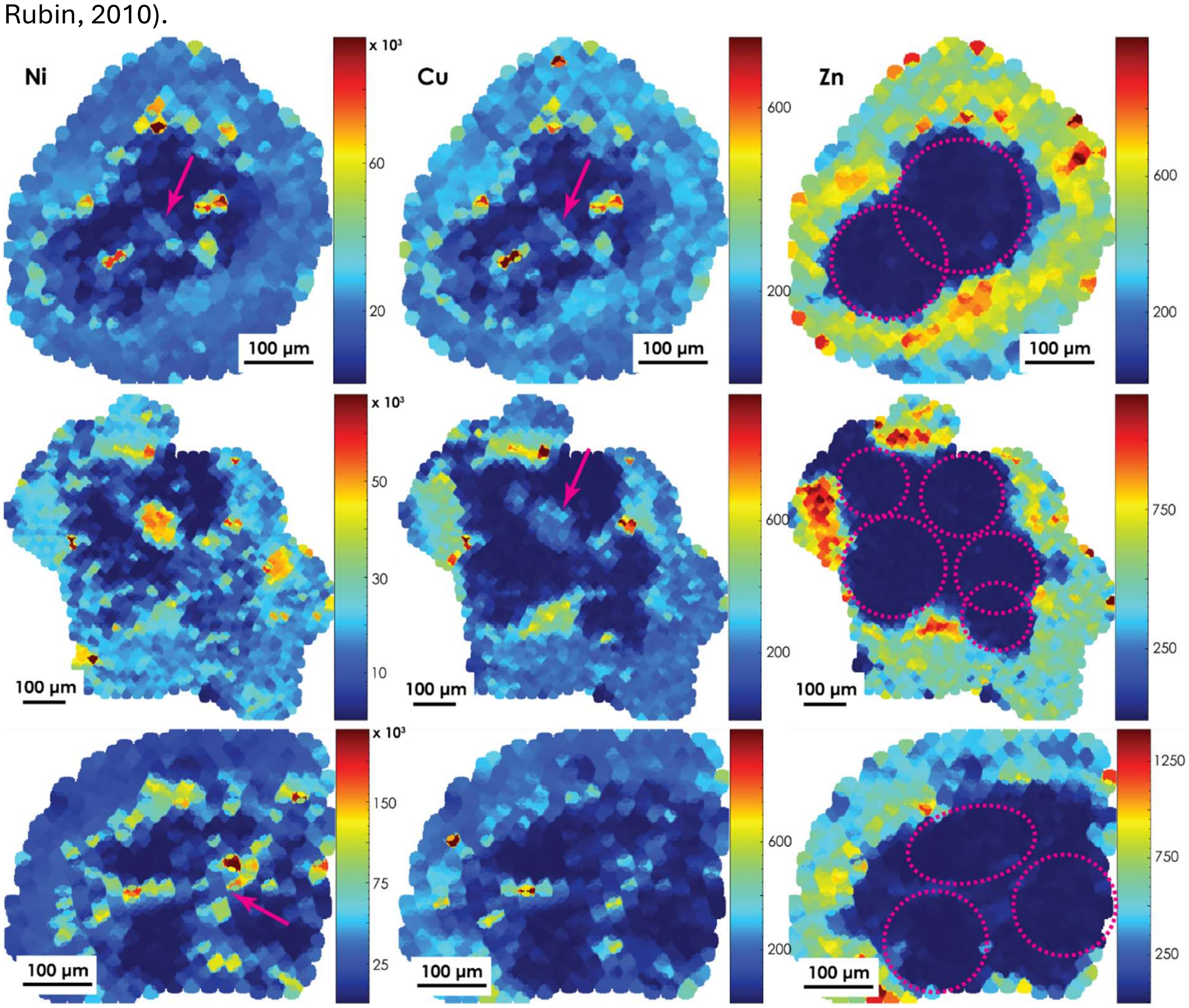

***Figure 10:*** *Ni, Cu and Zn elemental maps of three chondrules from Paris, that show microspherule regions partitioned by zones enriched in siderophile elements (i.e., Fe, Ni, Co, Cu; pink arrows). The color bars reflect µg/g concentrations of the elements. While in BSE mode and lithophile element space, these would appear as single ~300 µm diameter chondrules, they represent aggregates of ~100 µm grape-bunch chondrules (pink spheres), where these microchondrules contained metal-rich rims.*

Moreover, we observe no compositional differences between the polar, midlatitude and equatorial planes of chondrules (**Fig. 6C**), which is expected if we are sampling multiple smaller chondrules. These observations argue against CM microchondrules representing byproducts from the remelting of larger chondrules, as proposed for ordinary chondrite microchondrules (Krot et al., 1997; Bigolski et al., 2016). Instead, they imply that CM-like chondrules may, in many cases, represent aggregates of primary microchondrules that accreted while still hot or partially molten. In this view, microchondrules are not anomalous components, but fundamental building blocks of chondrules themselves.

This interpretation places constraints on the physical conditions of chondrule formation. Any viable environment must (1) melt dust only in small parcels, on the order of tens of micrometers; (2) cool melts rapidly enough to prevent immediate coalescence into larger droplets; and (3) maintain sufficiently high collision rates among hot particles to enable subsequent welding and aggregation into composite objects; (4) Chondrule formation must be highly localized and allow for preservation of fine-grained CI-like dust – including surviving complex organic matter and presolar grains – that rapidly accretes after chondrule formation. These conditions are difficult to reconcile with nebula-wide shock models, which typically predict spatially extensive heating and the direct formation of mm-scale molten droplets (Desch and Connolly Jr., 2002; Ciesla and Hood, 2002; Morris and Desch, 2010). Instead, they point toward localized, transient, and highly energetic processes operating within a dust-rich protoplanetary disk. Notably, many chondrule formation models aim to understand chondrule formation in the framework of mm-scale droplets and a 'microchondrules first' scenario has not been explored fully. While it is beyond the scope of this paper to frame our constraints into previously explored chondrule formation models, including for example current sheets (Joung et al., 2004; Hubbard et al., 2012; McNally et al., 2014) and nebular lightning (Horányi et al., 1995; Johansen and Okuzumi, 2018; Desch and Cuzzi, 2000), our observations suggest a shift in how chondrule formation is conceptualized. Rather than viewing chondrules as relatively large droplets that represent their initial dust aggregate diameter, they may instead represent mosaics of smaller droplets that survived, collided, and accreted within a dynamically violent, dust-rich disk. In this framework, chondrule formation becomes less a problem of one-shot melting and more a problem of granular physics operating under extreme, transient conditions. This perspective aligns naturally with dust-rich disk models, repeated heating events inferred from chondrule petrography and isotopic systematics (Hewins et al., 2005; Jones et al., 2018), and the pronounced textural complexity observed in chondrules.

*4.4 Chondrule formation and the MVE plateau*

In the model of Alexander (2019), calculated chondrule compositions do not exhibit an MVE plateau but instead show relatively continuous depletion as a function of 50% condensation temperature. Such behavior is expected for chondrule formation involving open-system evaporation within a canonical nebular environment, where elemental depletion follows relative volatility trends (Lodders, 2003). In contrast, Hellmann et al. (2020) derived an MVE plateau for their calculated non-matrix fraction, which they interpreted as reflecting bulk condensation from a CI-like gas reservoir. More recently, Zhang and Grewal (2026) have used updated matrix fractions from CM, CO and CR chondrites (Patzer et al., 2021, 2022, 2023) to recalculate the non-matrix fraction and came to similar conclusions as Alexander (2019): chondrules do not exhibit an MVE plateau This contrasts with our analyses of CM chondrules, which display an MVE plateau for elements with 50% condensation temperatures <1000 K. Petrographic observations provide important context for this feature. The grape-bunch microchondrule textures documented here show that individual "grapes" are locally separated by remnants of FGR material, which appear as compositional ghosts in the laser ablation data (**Fig. 11**). These textures indicate that bulk grape-bunch compositions include CI-like material in the form of primary FGRs.

***Figure 11:*** *Example of a Maribo main chondrule with two microchondrules with FGRs accreted to the main chondrule. The color bars reflect μg/g concentrations of the elements. The latter includes a Y-shaped region (pink arrows on Fe and Ga maps) that is roughly CI-like and not visible on the BSE image. This may be related to deeper sampling during LA-ICP-MS relative to BSE imaging.*

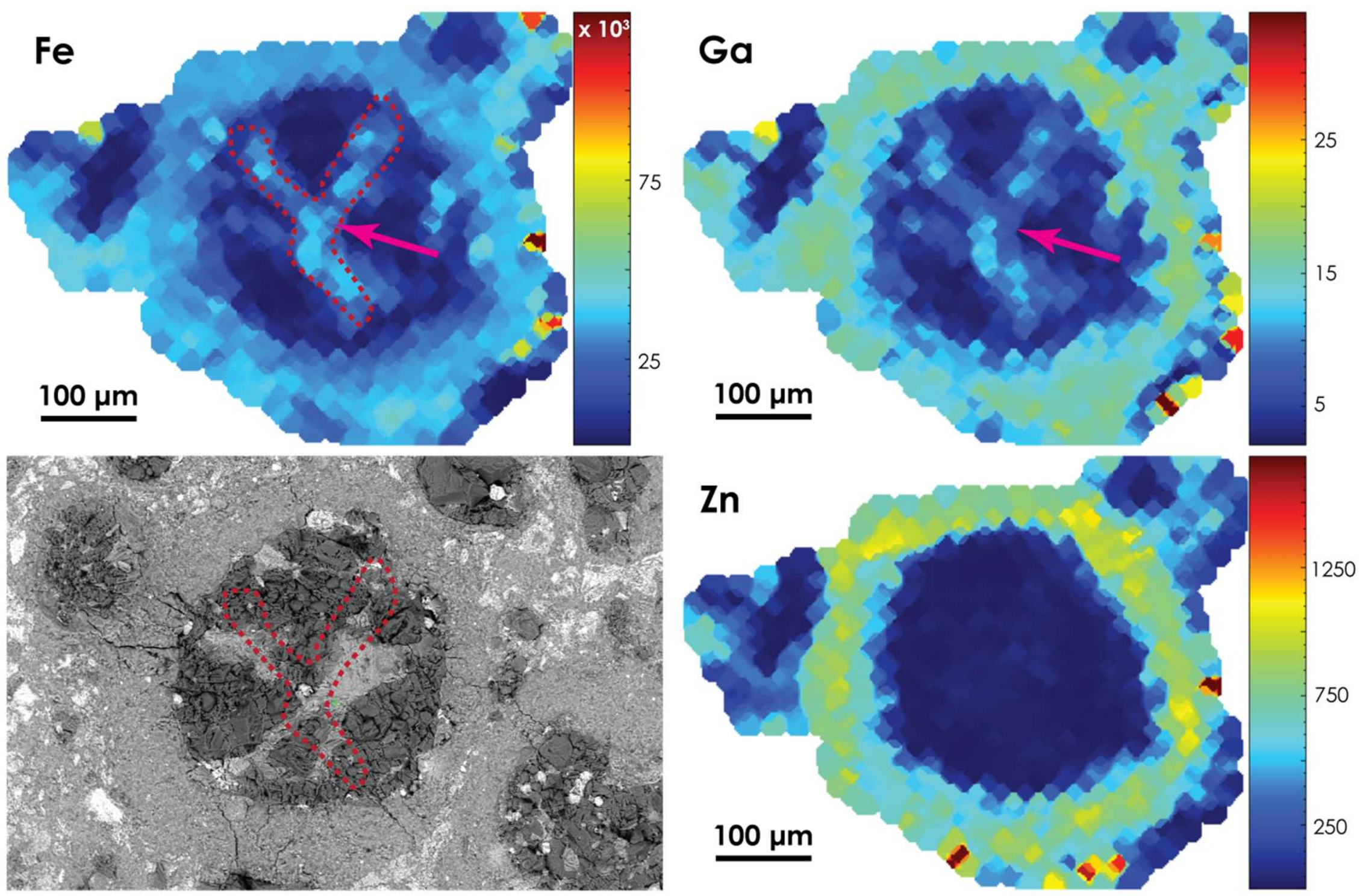

Although most of the initial rims were likely consumed during hot aggregation and partial melting of microchondrules into larger grape-bunch objects, in some cases their chemical signature would have been retained in these ghost rims (**Fig. 11**). This implies an addition of roughly 30 % of CI-like material to a canonical volatile-depleted chondrule composition to generate the observed MVE plateau at 0.29 × CI (**Fig. 4B**). This inferred CI fraction is higher than the ~13 % predicted by Hellmann, but our dataset is restricted to CM chondrules. These may have experienced distinct formation pathways relative to chondrules from other groups (e.g., CVs), consistent with isotopic differences reported by van Kooten et al. (2021, 2026).

These observations imply that MVE depletion in CM chondrules is not solely controlled by mesostasis evaporation, but also by the extent of accretion and incorporation of fine-grained CI-like dust. This dual control provides a natural explanation for the scatter observed in **Figure 8**, where MVE depletion is evaluated primarily as a function of evaporative loss, without explicitly accounting for variable dust addition.

If CI-like material was accreted onto microchondrules prior to their aggregation and welding, then the associated water ice could have modified the chondrules within the protoplanetary disk itself, rather than exclusively during parent-body alteration as represented by more altered CM chondrites where a progressive alteration sequence of chondrule and matrix phases is observed (CM < 2.6, Rubin et al., 2007). Such a scenario may explain why many CM chondrules display secondary alteration features and oxidation of metal and mesostasis, even though the host meteorites are considered among the most primitive CM specimens in current collections. This classification is largely based on the relatively limited alteration of the surrounding fine-grained rims and matrix, which preserve abundant GEMS-like materials. The apparent discrepancy suggests that chondrules may have undergone more extensive processing than their enclosing FGRs. Nevertheless, we recognize that even at the initial stages of parent body alteration, labile phases like glassy mesostasis, metal and melilite are rapidly degraded, accounting for the absence/paucity of the latter in CAIs from Paris and CM2.6 chondrite QUE 97990 (Rubin, 2007, 2015). We note, however, that melilite is more common in CAIs from Asuka 12236 (Kimura et al., 2020) – the least altered chondrite in our study – concomitant with a significant fraction of the chondrules showing alteration and oxidation of chondrules.
Accordingly, nebular alteration of CM chondrules should be considered as a viable and potentially significant process, consistent with models invoking aqueous or oxidizing conditions in the protoplanetary disk (Ciesla et al., 2003). This interpretation shifts part of the alteration history of CM chondrules from the parent body to the nebular environment and reinforces the importance of dust–chondrule interaction during outer disk evolution.

### *4.5 Chondrule formation in the outer Solar System*

Chondrules in carbonaceous chondrites record diverse formation environments across the Solar System (Tenner et al., 2015; Marrocchi et al., 2018), reflecting both spatial and temporal heterogeneity in chondrule-forming processes. In this study, we interpret CM chondrules as stereotypical of chondrule formation in the outer disk, since CM chondrites are most ubiquitous in our meteorite collection and many anomalous carbonaceous chondrites record similar texture and size distributions (van Kooten et al., 2026). Recently, microchondrules populations have also been recorded in Ryugu particles (Genge et al., 2025). However, not all carbonaceous chondrules share this origin, and several groups record fundamentally different formation pathways.

The clearest departures from disk-driven transient heating are observed in CH and CB chondrites, whose chondrules and metal-rich spherules are widely interpreted as products of high-energy impacts and formation within impact-generated melt plumes (Campbell et al., 2001; Sanders and Scott, 2012; Krot et al., 2021). Impact-related origins have also been proposed for CR chondrules, which formed relatively late (Schrader et al., 2017; Budde et al., 2018) and record Ni/Co ratios in chondrule metal consistent with metal–silicate equilibration at elevated pressures, which has been suggested to reflect formation from differentiated planetesimals (van Kooten et al., 2022). Similar considerations apply to the unusually large (mm–cm scale) chondrules in CV (and possibly CL) chondrites, whose nucleosynthetic isotopic systematics link to inner-disk reservoirs (van Kooten et al., 2021; Fukuda et al., 2024; Onyett et al., 2024). These large droplets are inconsistent with the small, composite chondrules typical of CM, CO, and anomalous carbonaceous chondrites and instead suggest derivation from distinct, high-temperature inner-disk environments, followed by radial transport. Future nucleosynthetic isotope studies of pristine CM-like chondrules would help us further understand their origin and formation history and their relationship to CV and CR chondrules that have been analyzed in multi-element isotope space owing to their relatively large sizes (Olsen et al., 2016; Gerber et al., 2017; Williams et al., 2020; Schneider et al., 2020; van Kooten et al., 2021).

Taken together, these lines of evidence suggest that only a subset of carbonaceous chondrites – most notably CM chondrites, along with CO and related anomalous carbonaceous chondrite groups – could preserve a record dominated by outer-disk, disk-driven transient heating. In this context, CM chondrules may represent the most faithful archive of canonical outer-disk chondrule formation, whereas chondrules in CH, CB, CR, and some CV and CL chondrites reflect significant contributions from impact-related processes or inner-disk sources that overprint or bypass the typical outer-disk environment.

## 5 Conclusions

We investigated the major, minor, and trace element compositions of 66 CM chondrules and associated fine-grained rims in relation to their three-dimensional morphology. These observations provide important constraints on the processes and mechanisms of chondrule formation in the outer protoplanetary disk.

First, CM chondrules record a systematic process of metal loss and evaporation of Si-rich mesostasis. This process drives initially CI-like chondrule precursor compositions toward more Mg- and Si-rich bulk compositions along the CI ratio line and ultimately toward more forsteritic compositions. The GEMS-like materials observed in pristine CM matrices appear to mirror the bulk compositions of chondrules and may represent complementary condensates derived from the Si-rich mesostasis vapor. The fine-grained dust accreted onto chondrules is dominantly CI-like but incorporates ~14 wt.% complementary condensate material represented by chondritic amorphous silicates in CM matrices. Together, these observations reconcile the Mg/Si complementarity between chondrules and matrix with the preservation of primordial organics and presolar grains in chondritic matrices.

Second, we find no evidence for significant sectioning bias in either chondrule size or sectioning plane. This may reflect the fact that CM chondrule size distributions are controlled by agglomerates of microspherules (~100 μm) rather than by larger primary molten droplets. Many chondrules in this study display grape-bunch textures composed of smaller primary chondrules that possessed metal-rich or CI-like rims prior to welding together. This observation may explain the moderately volatile element plateau at ~0.3×CI observed for the average CM chondrule composition, potentially reflecting the incorporation of primary FGR material into these grape-bunch aggregates, which exerts a strong control on the resulting MVE patterns.

Furthermore, water ice accreted onto these aggregates may have altered the chondrules through pervasive in situ oxidation and aqueous interaction, rather than exclusively through parent-body processes. Such alteration could have preferentially affected the welded chondrule interiors while leaving the surrounding FGRs comparatively less modified. Finally, this "micro-chondrule-first" scenario has important implications for chondrule formation models, as it suggests that chondrule formation may have occurred through highly localized thermal events that produced predominantly sub-100 μm molten droplets rather than larger melt bodies.

**Acknowledgements**
We thank Prof. James Day for efficient handling of our manuscript and Richard Ash, Alan Rubin and an anonymous referee, for insightful and detailed comments that have improved our manuscript. We thank the Natural History Museum of Paris and the Royal Belgian Institute of Natural Sciences for the loan of Paris section 4029sp3 (2010-3, B1.5) and Asuka 12236 section GEO758. We also thank Estrid Buhl Naver for assistance with micro-CT data reduction. The authors acknowledge funding from the Carlsberg Foundation (Semper Ardens: Advance grant FIRSTATMO) and from the Villum Foundation (Young Investigator grant nr. 53024 awarded to E.v.K.)

**Author contributions: CRediT**

Conceptualization: E.v.K., Methodology and analyses: P.E. and E.v.K., Funding acquisition: A.J., E.v.K., Writing – original draft: P.E., E.v.K., A.J.

**Data Availability**

All data presented in this study, including the numerical values underlying the figures and tables, are publicly available in the Mendeley Data repository at:

https://doi.org/10.17632/2wjcwcb69g.2

The repository includes all LA-ICP-MS datasets along with X-ray computed tomography scans of all samples.

**Appendix A. Supplementary Material**

The supplementary materials include BSE images and Mg-Fe-Si elemental maps of chondrules from Maribo, Paris and Asuka 12236, as well as Na and Al concentration core-rim profiles of representative chondrules from Maribo and Paris.

Supplementary excel sheet: includes individual bulk chondrule and FGR compositional data of the 66 chondrules analyzed by LA-ICP-MS in this study.

Supplementary videos: includes low and high resolution µCT scans of Maribo, Paris and Asuka 12236.

| | Maribo (A) | | | | Maribo (B) | | | | Paris | | | | Asuka 12236 | | | | Average CM | | | |
|---|---|---|---|---|---|---|---|---|---|---|---|---|---|---|---|---|---|---|---|---|
| | rim | 2σ | core | 2σ | rim | 2σ | core | 2σ | rim | 2σ | core | 2σ | rim | 2σ | core | 2σ | rim | 2σ | core | 2σ |
| **Na** | 2294 | 168 | 818 | 186 | 2505 | 307 | 1593 | 460 | *6257* | *332* | 1758 | 394 | 2960 | 368 | 1296 | 771 | 3504 | 1856 | 1366 | 413 |
| **Mg** | 111779 | 9001 | 205229 | 26147 | 108067 | 4211 | 231424 | 66806 | 104542 | 3711 | 210784 | 21641 | 109831 | 5447 | 172296 | 22362 | 108555 | 3075 | 204933 | 24503 |
| **Al** | 15980 | 1417 | 10974 | 2820 | 15280 | 1109 | 10762 | 3107 | 14361 | 647 | 10569 | 2963 | 12531 | 5352 | 9727 | 1707 | 14538 | 1493 | 10508 | 546 |
| **Si** | 154021 | 11208 | 195033 | 16773 | 144760 | 18881 | 201072 | 201072 | 135585 | 19559 | 176971 | 58376 | 145465 | 4514 | 162836 | 12033 | 144958 | 7534 | 183978 | 17421 |
| **P** | 2364 | 392 | 681 | 243 | 2243 | 224 | 799 | 231 | 1503 | 93 | 963 | 324 | 1515 | 105 | 980 | 278 | 1906 | 461 | 856 | 142 |
| **S** | n.a. | | | | 17259 | 5444 | 4822 | 1392 | 27288 | 5795 | 4221 | 655 | n.a. | | | | 17259 | 10030 | 4522 | 600 |
| **K** | *336* | *45* | 116 | 21 | *335* | *102* | 165 | 48 | 639 | 87 | 116 | 22 | 689 | 110 | 151 | 78 | 664 | 50 | 137 | 25 |
| **Ca** | 7898 | 930 | 13219 | 6318 | 7395 | 383 | 8783 | 2535 | 11340 | 1599 | 9185 | 3098 | *20031* | *3207* | *26273* | *4823* | 11666 | 2479 | 10396 | 2833 |
| **Ti** | 587 | 60 | 1515 | 663 | 557 | 35 | 1127 | 325 | 581 | 30 | 802 | 174 | 490 | 295 | 605 | 106 | 554 | 45 | 1012 | 398 |
| **V** | 68 | 6 | 134 | 52 | 64 | 6 | 149 | 43 | 63 | 2 | 99 | 9 | 58 | 5 | 79 | 11 | 63 | 4 | 115 | 32 |
| **Cr** | 5326 | 421 | 4975 | 517 | 5645 | 329 | 5072 | 1464 | 4379 | 246 | 5208 | 672 | 4281 | 186 | 4503 | 575 | 4908 | 681 | 4939 | 306 |
| **Mn** | 2673 | 146 | 2033 | 448 | 2897 | 179 | 1630 | 470 | 2687 | 101 | 1776 | 427 | 2815 | 191 | 1693 | 524 | 2768 | 107 | 1783 | 177 |
| **Fe** | 318295 | 27057 | 139710 | 58006 | 288336 | 14011 | 94656 | 27325 | 344064 | 17984 | 163336 | 41979 | 316223 | 11255 | 219597 | 47529 | 316729 | 22775 | 154325 | 52012 |
| **Co** | 1212 | 85 | 138 | 27 | 1202 | 129 | 163 | 47 | 864 | 44 | 344 | 109 | 720 | 52 | 489 | 157 | 999 | 246 | 284 | 165 |
| **Ni** | 26508 | 1193 | 2891 | 418 | 26138 | 2030 | 4115 | 1188 | 21199 | 1213 | 8100 | 2677 | 19192 | 1245 | 11197 | 3949 | 23259 | 3634 | 6576 | 3800 |
| **Cu** | 273 | 30 | 45 | 18 | 267 | 29 | 66 | 19 | 215 | 10 | 48 | 17 | 200 | 14 | 68 | 28 | 239 | 37 | 57 | 12 |
| **Zn** | 718 | 104 | 17.7 | 7.1 | 723 | 82 | *60* | *17* | 515 | 34 | 12.2 | 4.6 | 595 | 55 | 25.1 | 9.6 | 638 | 101 | 18.3 | 7.5 |
| **Ga** | 13.4 | 1.0 | 2.3 | 1.3 | 13.6 | 0.7 | 3.2 | 0.9 | 13.2 | 0.8 | 2.4 | 0.7 | 15.1 | 1.4 | 4.2 | 1.3 | 13.8 | 0.9 | 3.1 | 1.2 |
| **Ge** | 45.6 | 5.0 | 11.2 | 2.9 | 43.6 | 3.5 | 12.1 | 3.5 | 40.3 | 2.6 | 6.2 | 1.8 | 50.0 | 7.2 | 21.5 | 12.7 | 44.9 | 4.1 | 12.8 | 9.0 |
| **Se** | 47.2 | 8.8 | 5.8 | 1.2 | 28.4 | 2.7 | 5.9 | 1.7 | 22.2 | 1.4 | 4.2 | 1.9 | 25.4 | 2.9 | 4.0 | 3.0 | 30.8 | 11.2 | 5.0 | 1.1 |
| **Rb** | 2.7 | 0.3 | 0.5 | 0.2 | 2.9 | 0.6 | 0.9 | 0.3 | 2.9 | 0.4 | 0.5 | 0.2 | 4.0 | 0.5 | 0.6 | 0.3 | 3.1 | 0.6 | 0.6 | 0.1 |
| **Zr** | 7.2 | 0.9 | 11.2 | 5.3 | 7.0 | 0.6 | 8.1 | 2.3 | 5.3 | 0.3 | 5.6 | 1.1 | 4.9 | 2.0 | 5.8 | 1.1 | 6.1 | 1.2 | 7.7 | 3.7 |
| **Mo** | 1.66 | 0.42 | 0.78 | 0.47 | 1.62 | 0.38 | 0.81 | 0.23 | 1.64 | 0.11 | 0.92 | 0.21 | 1.30 | 0.18 | 0.88 | 0.27 | 1.55 | 0.17 | 0.85 | 0.08 |
| **Ag** | 0.36 | 0.04 | 0.05 | 0.01 | 0.34 | 0.04 | 0.05 | 0.01 | 0.29 | 0.02 | 0.09 | 0.02 | b.d. | | | | 0.33 | 0.04 | 0.06 | 0.02 |
| **Cd** | 1.87 | 0.27 | 0.04 | 0.01 | 2.09 | 0.28 | 0.17 | 0.05 | 1.41 | 0.15 | 0.07 | 0.02 | 1.88 | 0.22 | 0.18 | 0.06 | 1.81 | 0.29 | 0.12 | 0.09 |
| **Sn** | 1.72 | 0.19 | 0.42 | 0.10 | 1.66 | 0.14 | 0.47 | 0.14 | 1.47 | 0.12 | 0.24 | 0.05 | *2.51* | *0.35* | *1.10* | *0.36* | 1.84 | 0.46 | 0.56 | 0.52 |
| **Te** | 4.07 | 0.48 | 0.50 | 0.16 | 3.73 | 0.35 | 0.62 | 0.18 | 2.64 | 0.27 | 0.53 | 0.22 | 3.30 | 0.42 | 0.56 | 0.26 | 3.44 | 0.62 | 0.55 | 0.04 |
| **La** | n.a. | | | | 0.56 | 0.09 | 0.34 | 0.10 | 0.37 | 0.06 | 0.40 | 0.21 | n.a. | | | | 0.47 | 0.20 | 0.37 | 0.05 |
| **W** | 0.27 | 0.10 | 0.10 | 0.07 | 0.30 | 0.06 | 0.09 | 0.03 | 0.16 | 0.01 | 0.08 | 0.03 | b.d. | | | | 0.24 | 0.07 | 0.09 | 0.01 |

| | | | | | | | | | | | | | | | | |
|---|---|---|---|---|---|---|---|---|---|---|---|---|---|---|---|---|
| **Au** | 0.27 | 0.04 | 0.04 | 0.01 | 0.25 | 0.02 | 0.06 | 0.02 | 0.31 | 0.03 | 0.13 | 0.03 | *0.78* | *0.24* | *0.85* | *0.34* | 0.28 | 0.04 | 0.08 | 0.05 |
| **Pb** | 3.83 | 0.67 | 0.68 | 0.29 | 3.80 | 0.35 | 0.95 | 0.27 | 3.22 | 0.30 | 0.43 | 0.19 | *7.62* | *0.92* | *4.58* | *0.64* | 3.62 | 0.40 | 0.69 | 0.30 |

***Table 2**: Averaged rim and chondrule (core) compositions in µg/g with 2σ uncertainty of Maribo (section A and B), Paris and Asuka 12236, as well as the total average of all analyzed chondrules and rims (n=66). N.a. = not analyzed, b.d. = below detection limit. Italic values are not included in the averaged CM chondrule and rim composition.*

# Supplementary Materials

## Chondrule formation in the outer disk from the primary 3D chemical composition of CM chondrules

Eyðbjørnsdóttir et al.

Fig. S1-S10

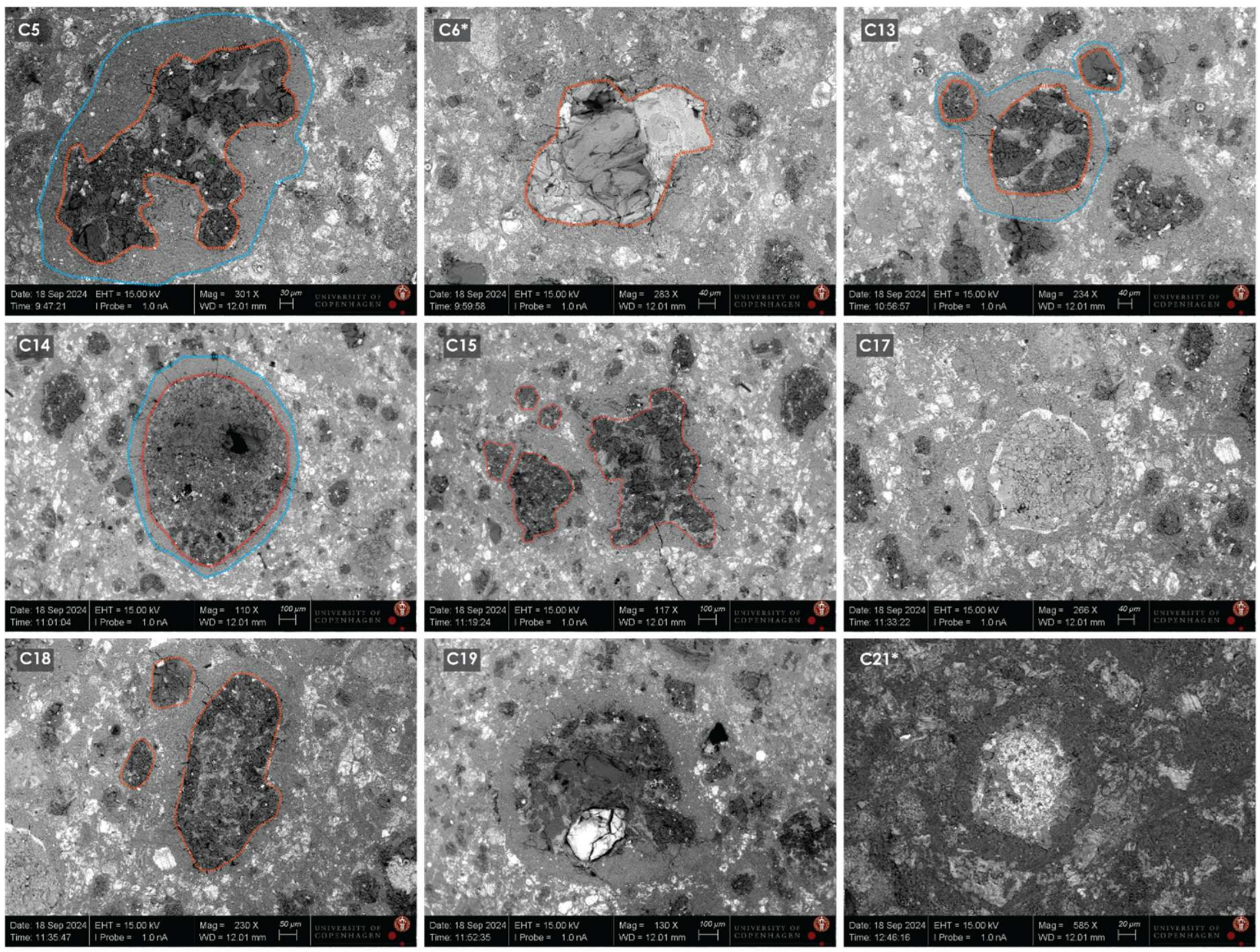

***Fig. S1:*** *BSE images of chondrules (orange) with their fine-grained rims (blue) from Maribo section A used in this study. Chondrules with * are not included in the discussion of the main text.*

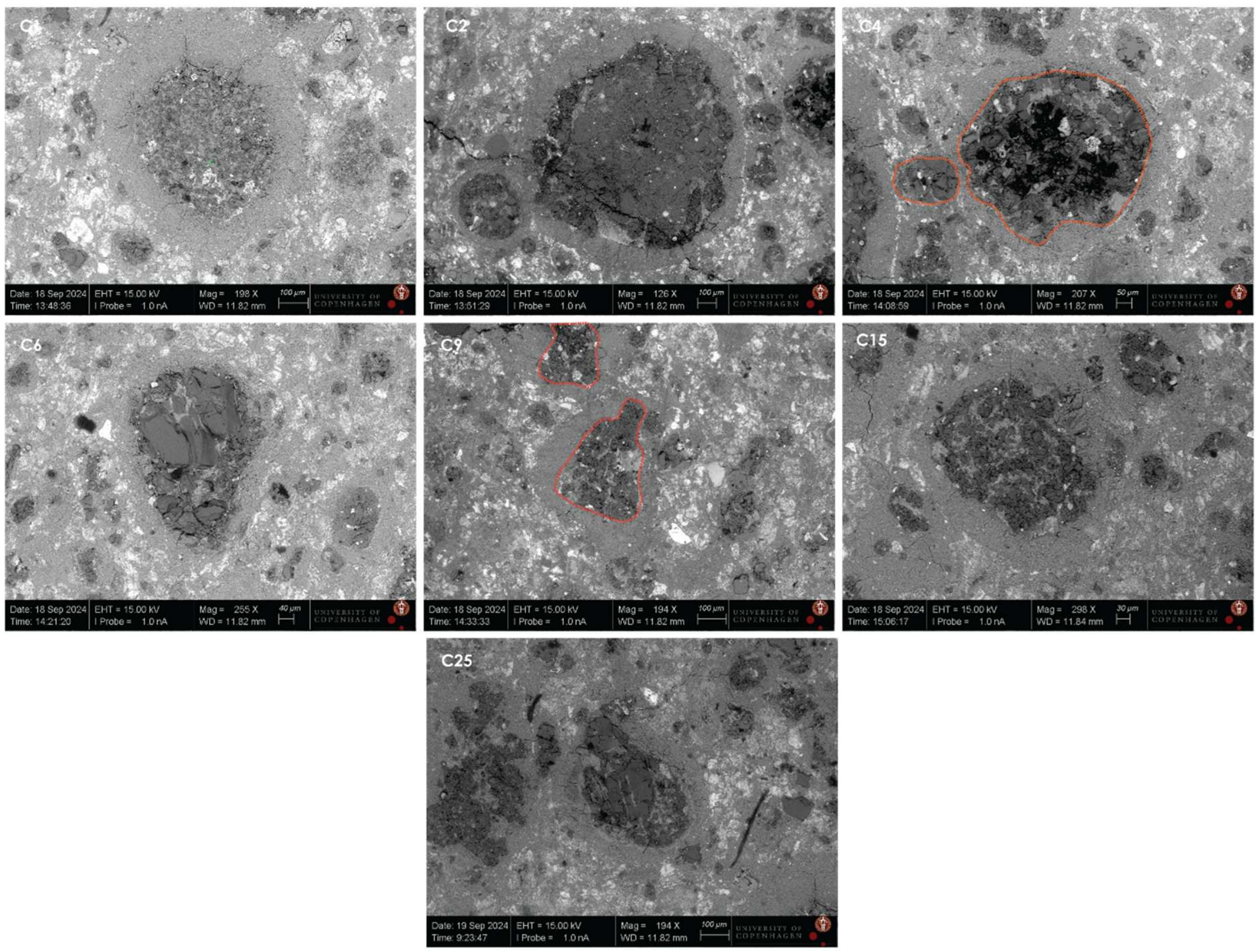

***Fig. S2:*** *BSE images of chondrules (orange) with their fine-grained rims from Maribo section B used in this study.*

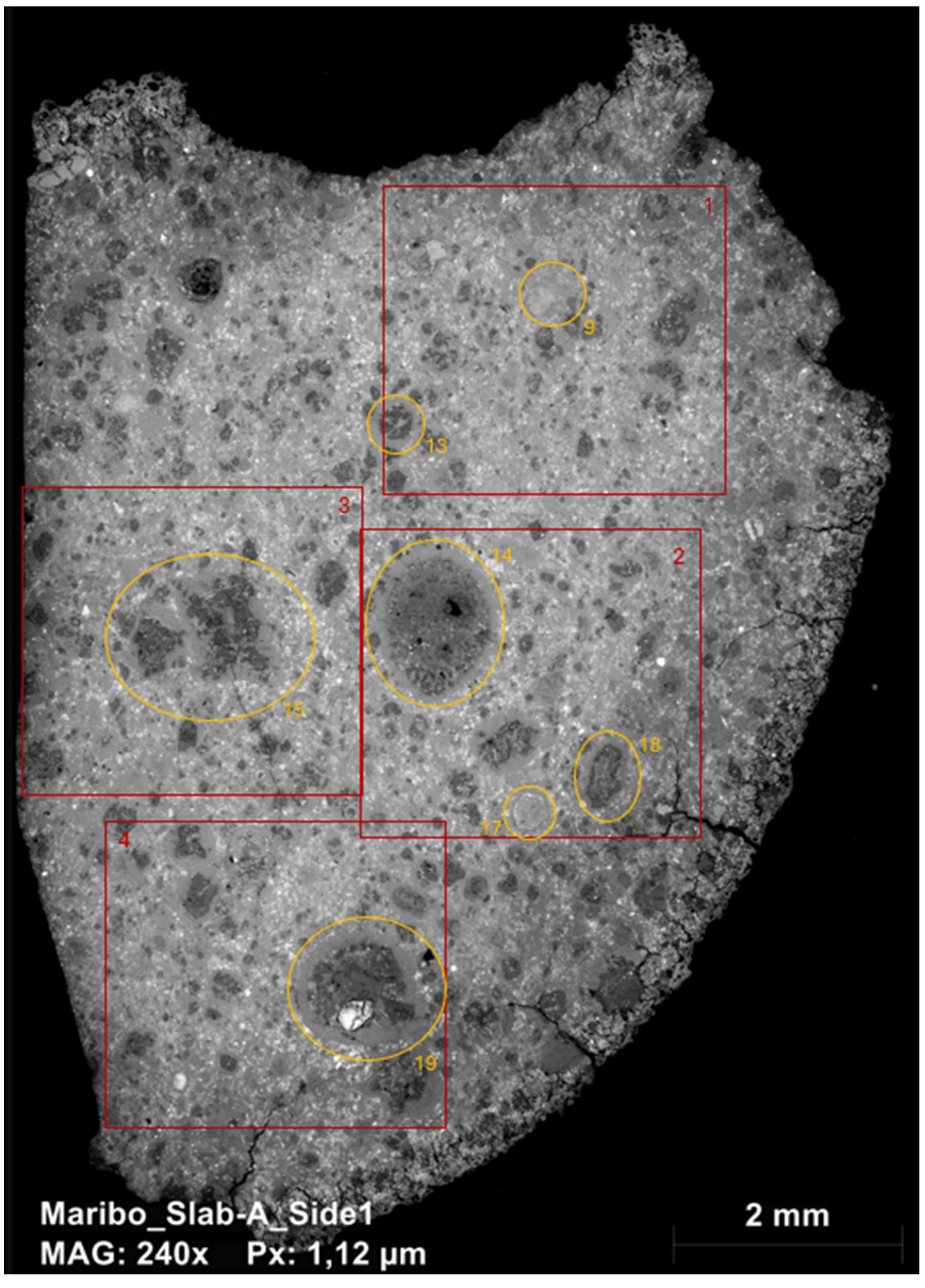

***Figure S3:*** *Maribo section A. Red boxes show the areas that were scanned in high resolution mode using micro-CT. Yellow circles show the chondrules studied in this paper.*

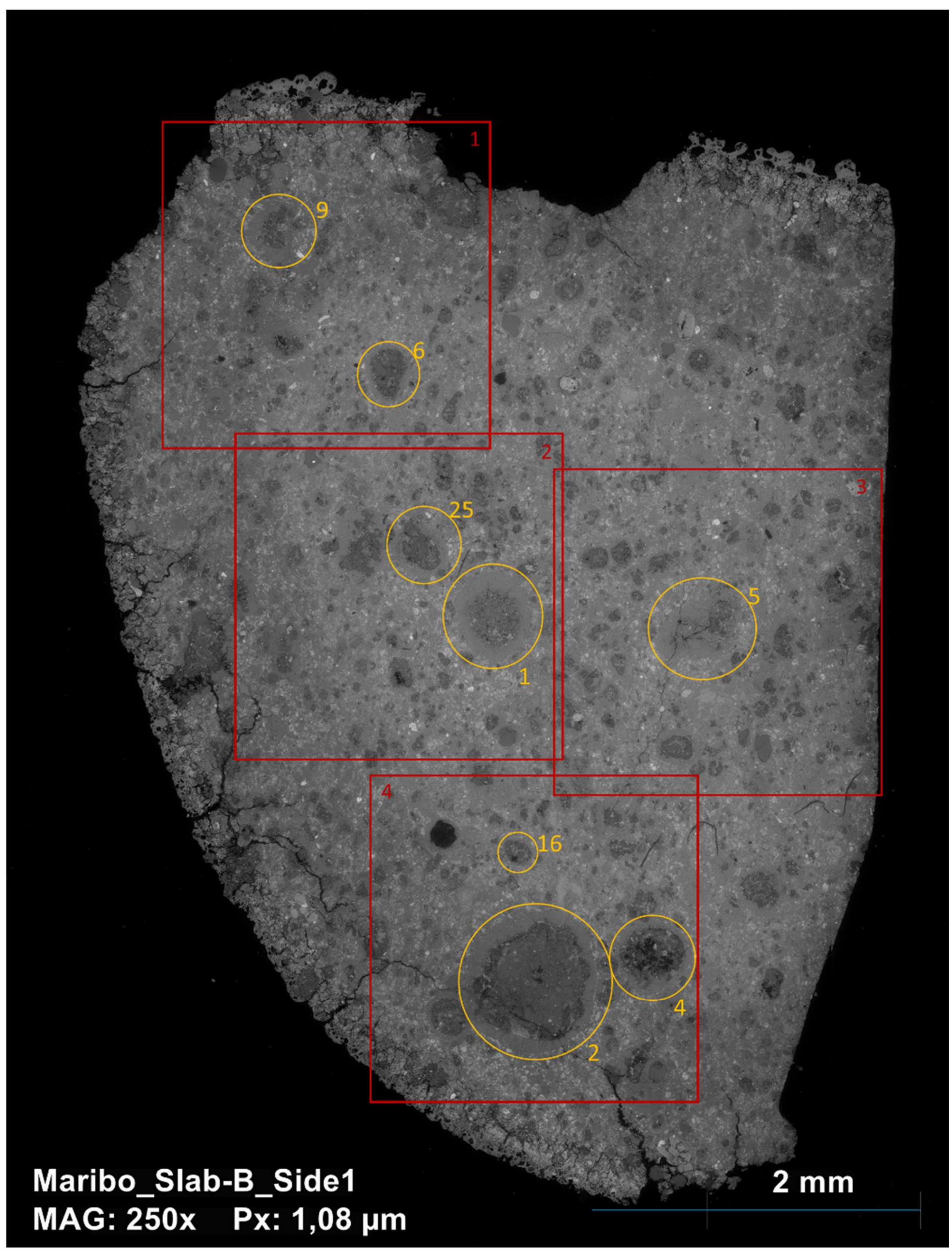

***Figure S4:*** *Maribo section B. Red boxes show the areas that were scanned in high resolution mode using micro-CT. Yellow circles show the chondrules studied in this paper.*

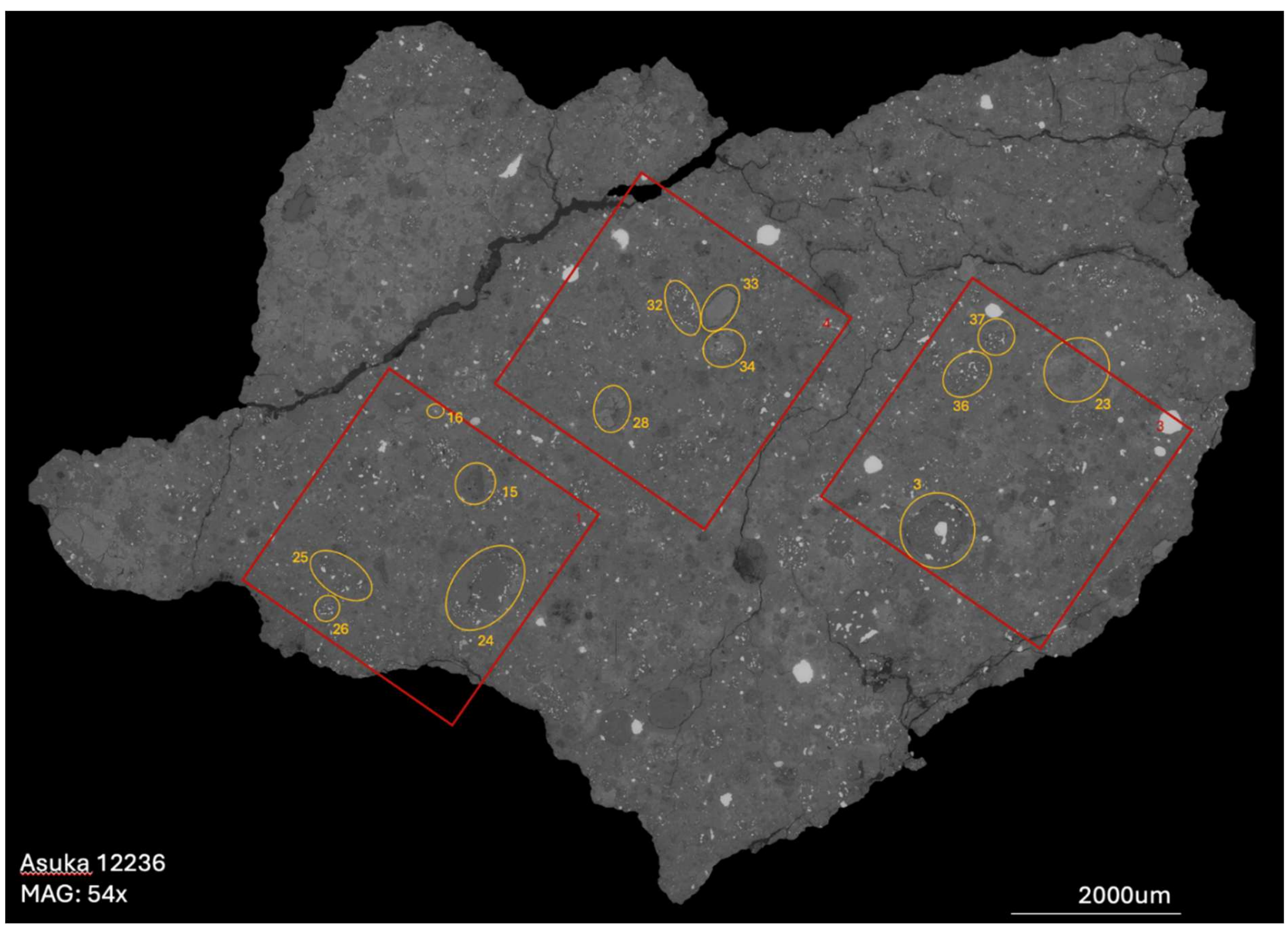

***Figure S5:*** *Asuka 12236. Red boxes show the areas that were scanned in high resolution mode using micro-CT. Yellow circles show the chondrules studied in this paper.*

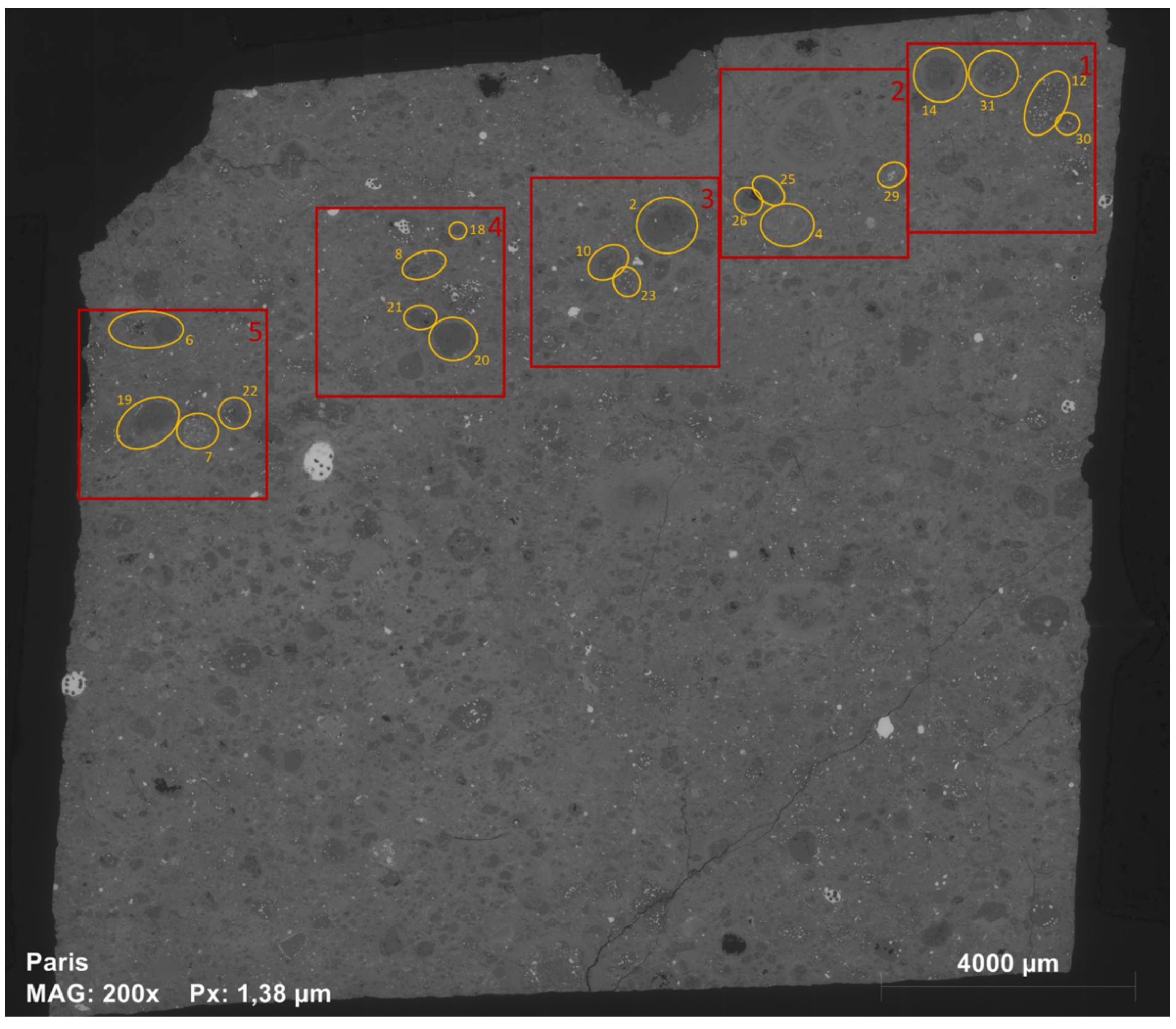

***Figure S6:*** *Paris. Red boxes show the areas that were scanned in high resolution mode using micro-CT. Yellow circles show the chondrules studied in this paper.*

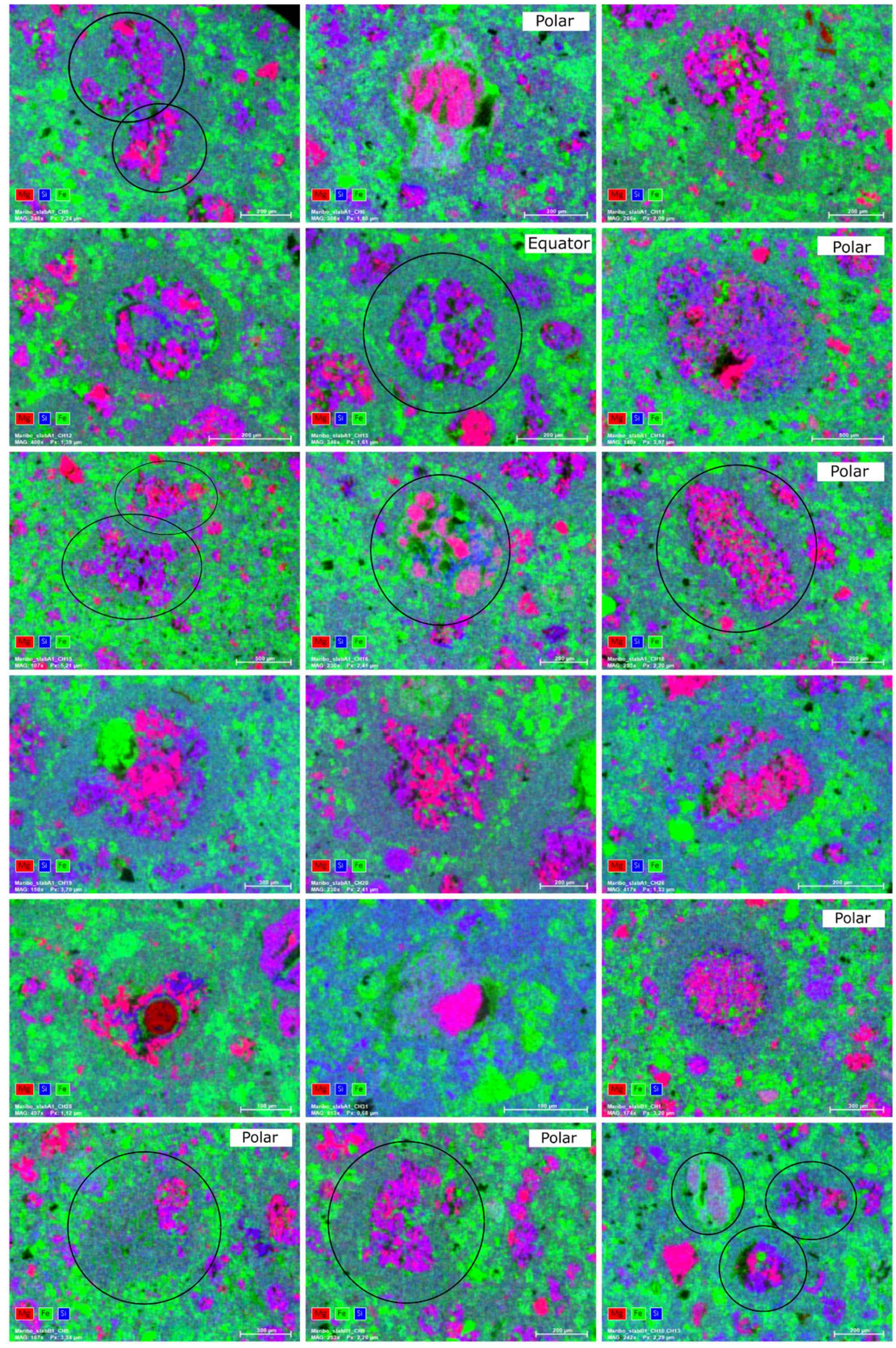

***Figure S7:*** *Mg-Fe-Si (RGB) elemental maps of chondrules from section A and B of Maribo, including chondrules for which their sectioning plane has been determined by X-ray tomography and have been analyzed using LA-ICPMS. Pink colors indicate forsterite, whereas purple indicates low-Ca pyroxene. These chondrules show limited evidence of low-Ca pyroxene zonation enveloping forsterite-rich cores as described by Friend et al. (2016).*

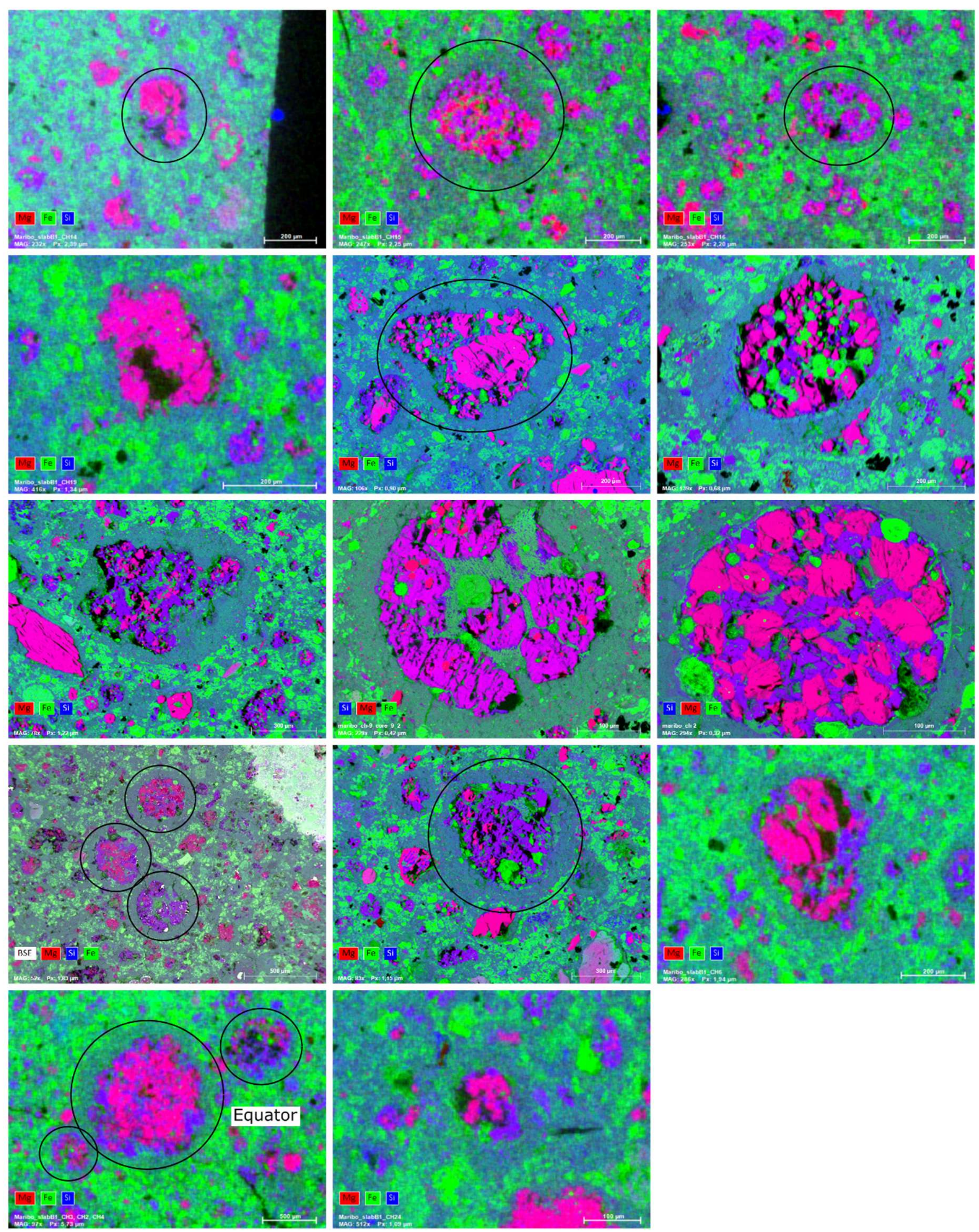

***Figure S7:*** *continued.*

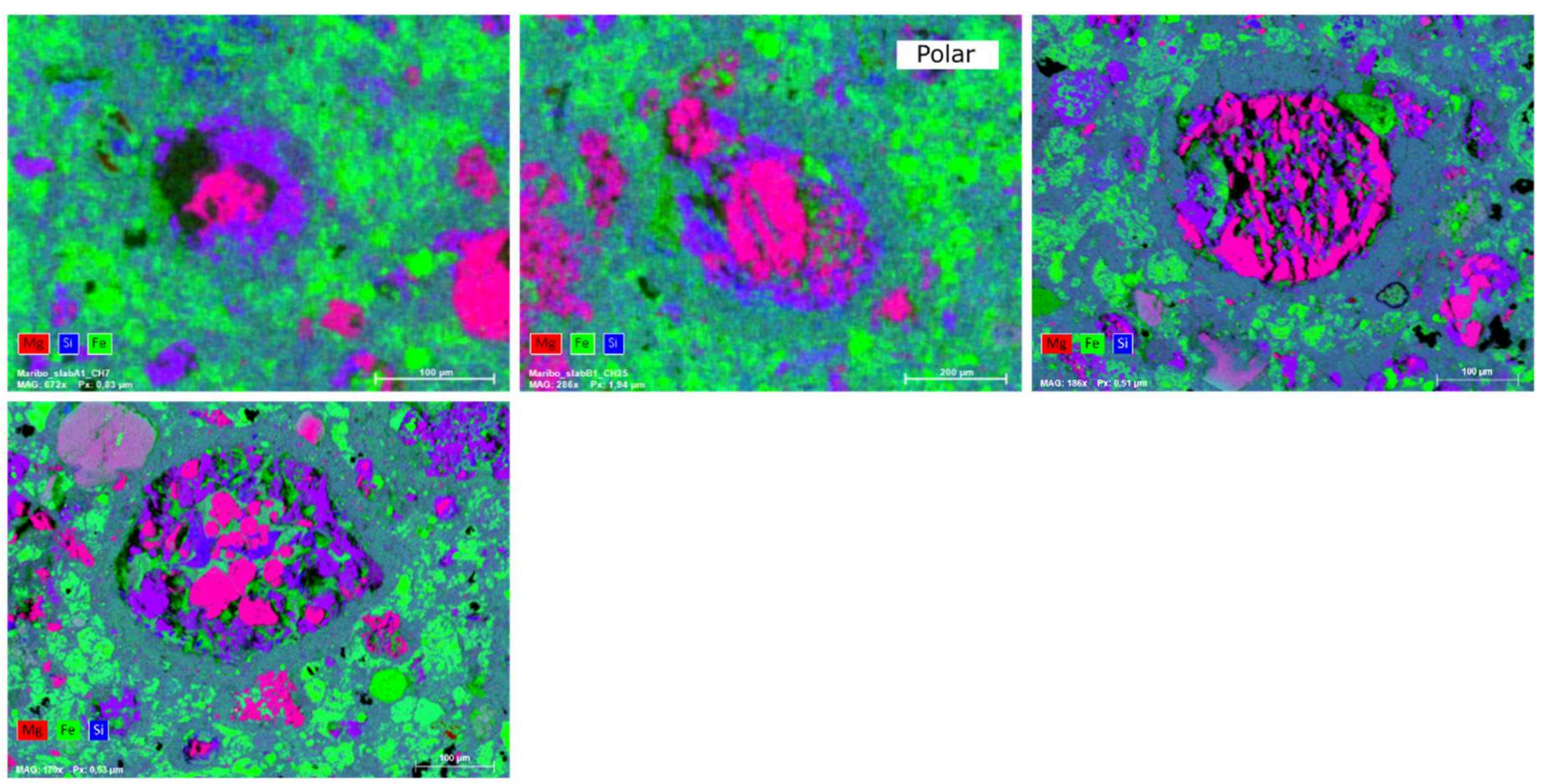

***Figure S7:*** *continued. Maribo chondrules showing evidence of low-Ca pyroxene zonation enveloping a forsterite chondrule core.*

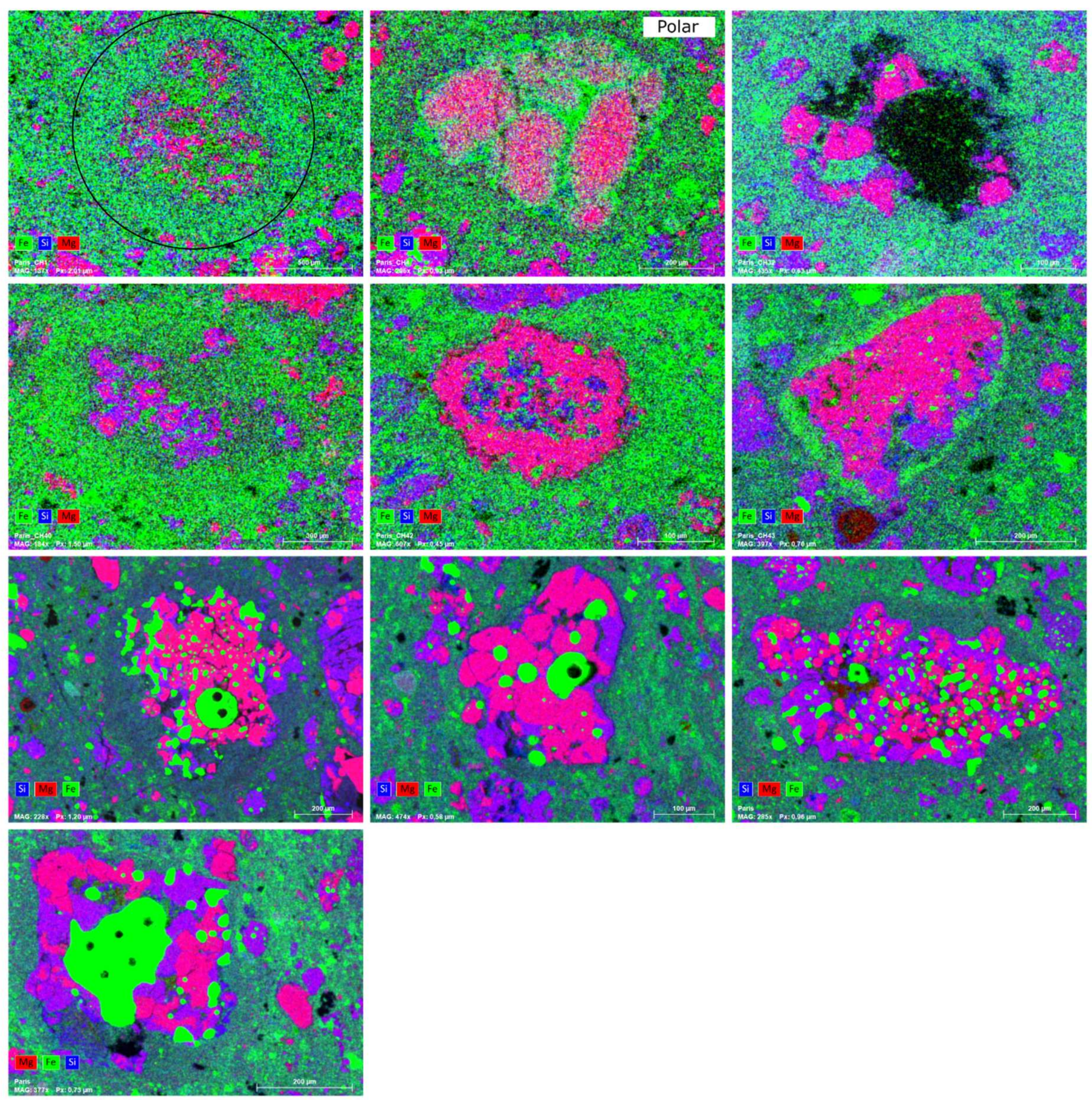

***Figure S8:*** *Mg-Fe-Si (RGB) elemental maps of chondrules from Paris, including chondrules for which their sectioning plane has been determined by X-ray tomography and have been analyzed using LA-ICPMS. Pink colors indicate forsterite, whereas purple indicates low-Ca pyroxene.*

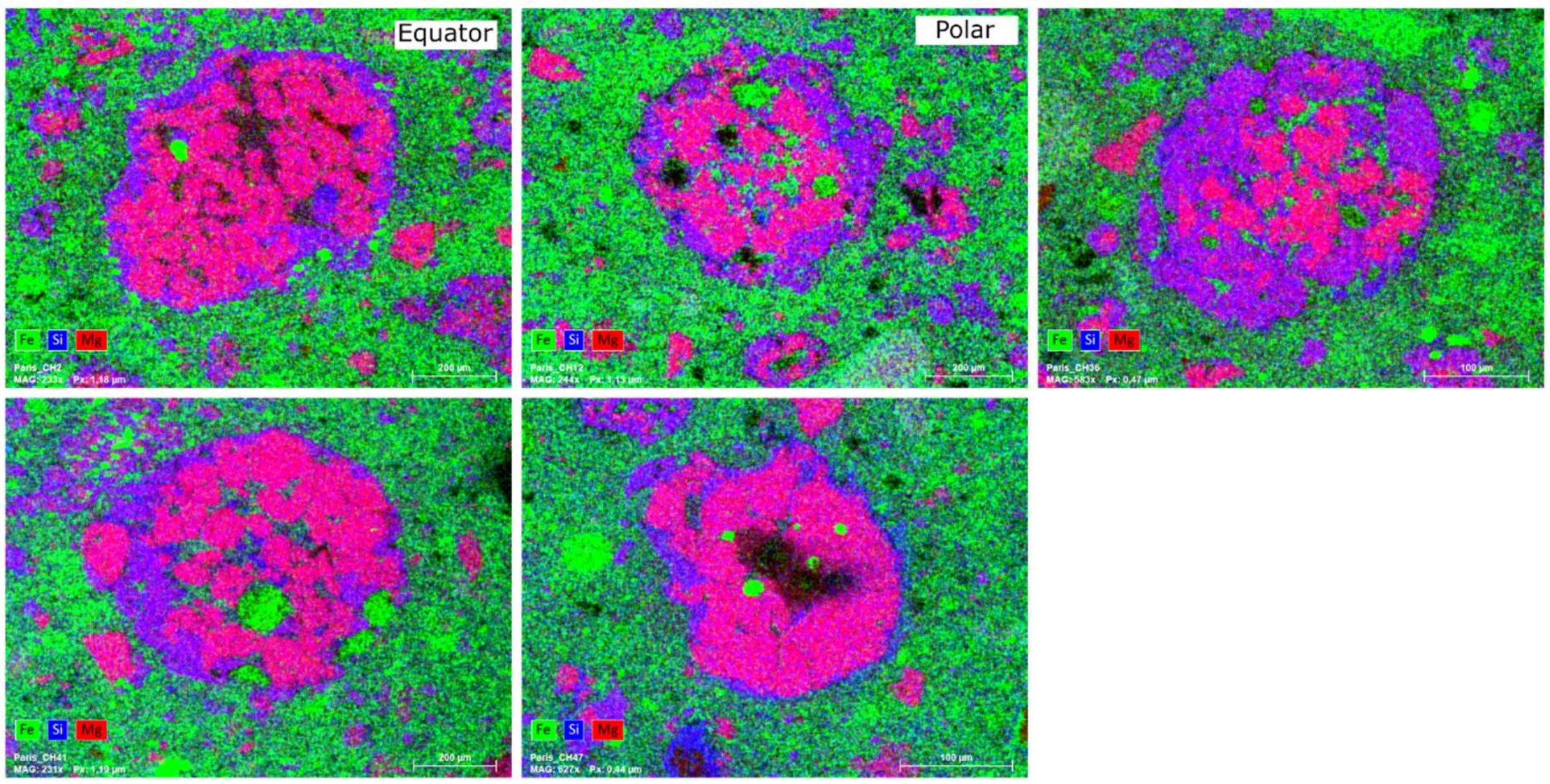

***Figure S8:*** *continued. Paris chondrules showing evidence of low-Ca pyroxene zonation enveloping a forsterite chondrule core.*

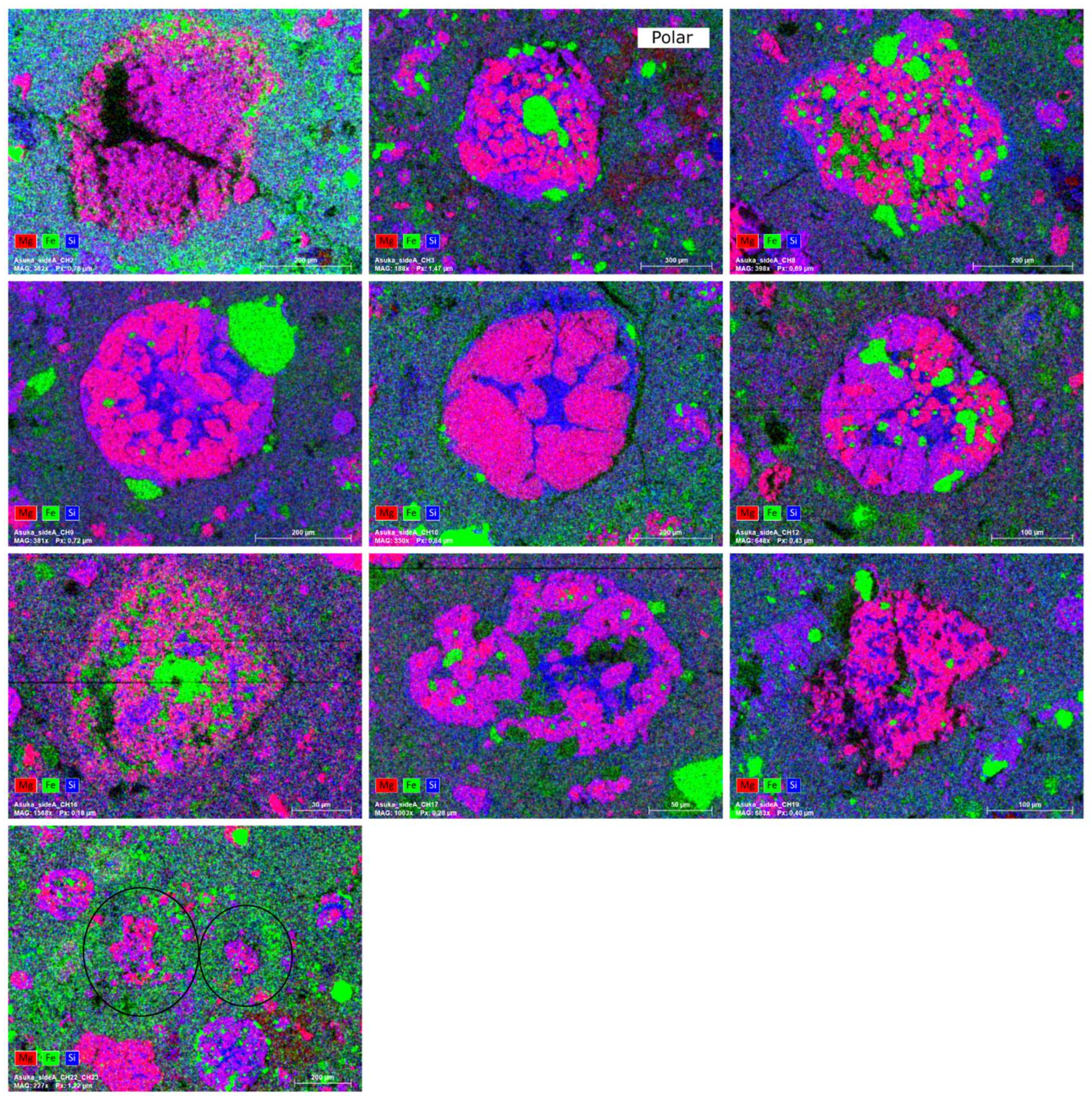

***Figure S9:*** *Mg-Fe-Si (RGB) elemental maps of chondrules from Asuka 12236, including chondrules for which their sectioning plane has been determined by X-ray tomography and have been analyzed using LA-ICPMS. Pink colors indicate forsterite, whereas purple indicates low-Ca pyroxene.*

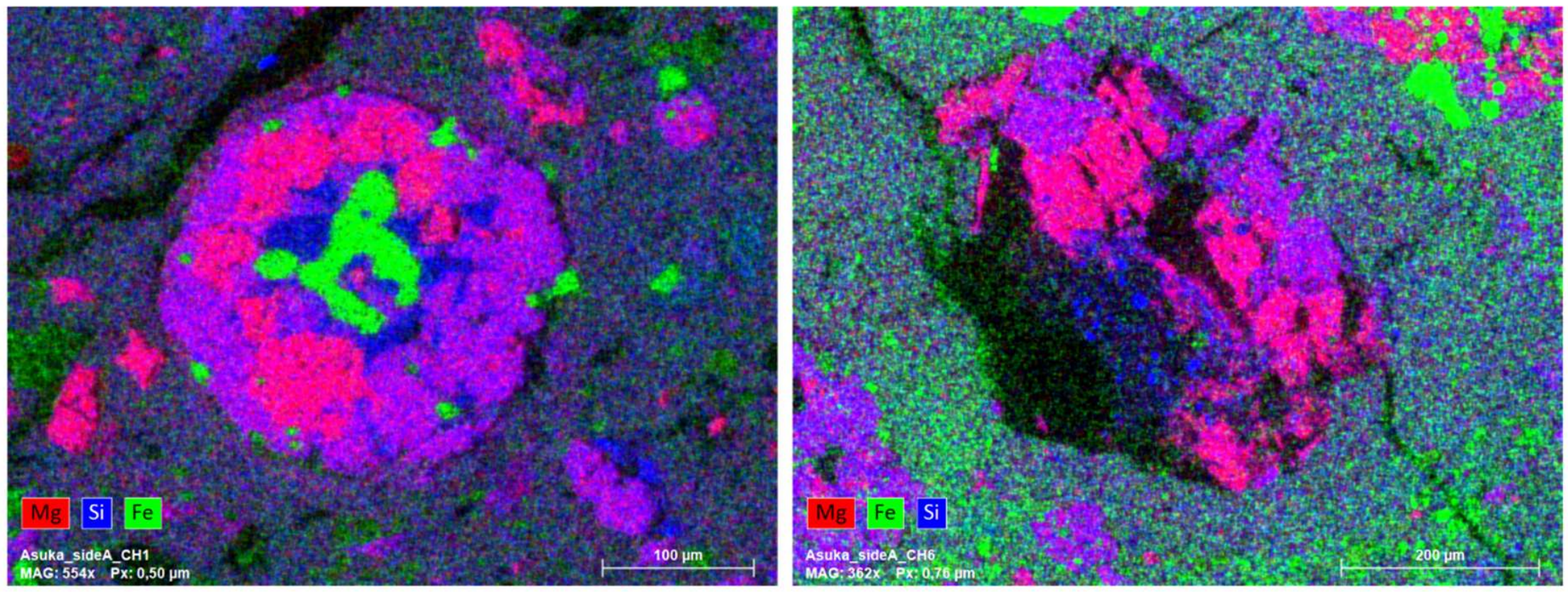

***Figure S9:*** *continued. Asuka 12236 chondrules showing evidence of low-Ca pyroxene zonation enveloping a forsterite chondrule core.*

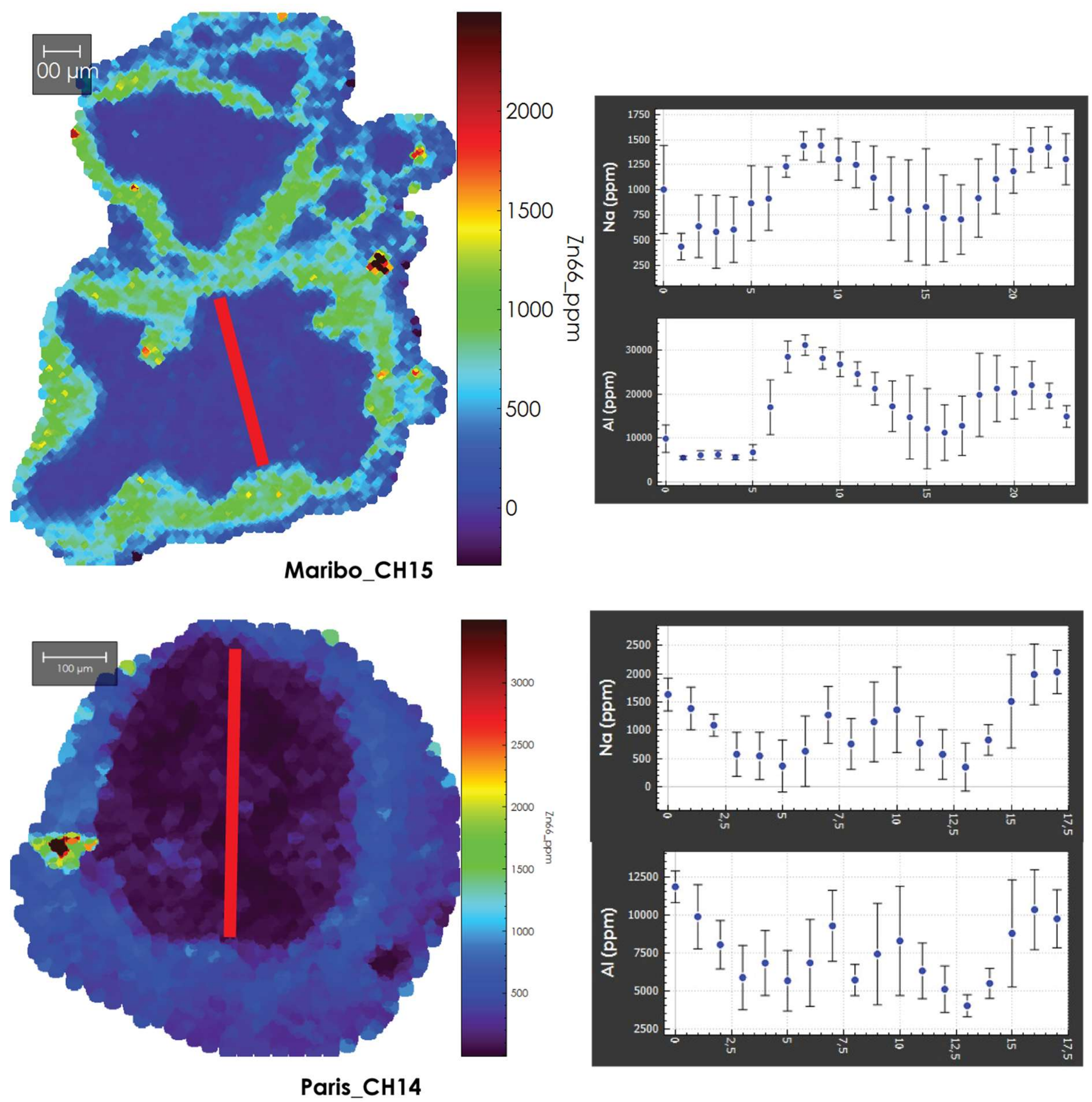

***Figure S10:*** *Laser ablation ICPMS obtained Zn elemental maps of Maribo CH15 and Paris CH14 chondrules and their fine-grained rims (Zn-rich), which are representative of the overall chondrule dataset in this study. Profiles (red lines) through these chondrules are spaced in 20 × 50 µm (length × width) boxes with Na and Al contents in ppm shown in the panels on the right. Note that the Na variation follows that of Al, suggesting that this variation is related to presence of mesostasis.*